\newcommand{\nc}{\newcommand}
\nc{\la}{\lambda} \nc{\alf}{\alpha}
\nc{\tht}{\theta}  \nc{\be}{\beta}  \nc{\eps}{\epsilon}
\nc{\ga}{\gamma}  \nc{\Ga}{\Gamma}
\nc{\de}{\delta} \nc{\si}{\sigma}  \nc{\ka}{\kappa}
\nc{\om}{\omega}  \nc{\qq}{\quad\quad}
\nc{\nf}{\infty}   \nc{\dl}{\mathop{\smash{\cal L}}}
\nc{\ra}{\rightarrow}  \nc{\ol}{\overline}
\nc{\beq}{\begin{equation}}  \nc{\barr}{\begin{array}}
\nc{\earr}{\end{array}}
\nc{\eeq}{\end{equation}}
\nc{\beqa}{\begin{eqnarray}}  \nc{\dst}{\displaystyle}
\nc{\eeqa}{\end{eqnarray}} \nc{\nnb}{\nonumber}
\nc{\bs}{\backslash}        \nc{\mbb}{\mathbb}
\nc{\brm}{\begin{remunerate}} \nc{\erm}{\end{remunerate}}
\nc{\nn}{\nonumber \\}
\nc{\p}[1]{{(\ref{#1})}}
\newcounter{muni}
{\usecounter{muni}
\setlength{\leftmargin}{0pt}\setlength{\itemindent}{38pt}

\newcommand\T{\theta_{12}}
\newcommand\Tb{{\bar\theta}_{12}}
\newcommand\Z{Z_{12}}
\newcommand\D{{\cal D}}
\newcommand\Db{\overline{\cal D}}
\newenvironment{remunerate}{\begin{list}{{\rm \arabic{muni}.}}
{\usecounter{muni}
\setlength{\leftmargin}{0pt}\setlength{\itemindent}{38pt}}}{\end{list}}
\documentclass[12pt]{article}  
\usepackage{amssymb}
\renewcommand{\thefootnote}{\fnsymbol{footnote}}
\def\theequation{\arabic{section}.\arabic{equation}}
\topmargin=-2cm\textheight=23.5cm\textwidth=16.cm
\oddsidemargin=0.25cm\evensidemargin=0.25cm
\begin{document}
\begin{titlepage}
\begin{flushright}
LPTHE 99-47 \\
JINR E2-99-340 \\
hep-th/0001165 \\
December 1999
\end{flushright}
\vskip 1.0truecm
\begin{center}
{\large \bf Quaternionic Metrics From Harmonic Superspace:\\
Lagrangian Approach and Quotient Construction}
\end{center}
\vskip 1.0truecm
\centerline{\bf Evgeny Ivanov${}^{\;a,1}$, Galliano Valent${}^{\;b,2}$}
\vskip 1.0truecm
\centerline{${}^{a}$\it Bogoliubov Laboratory of Theoretical
Physics, JINR,}
\centerline{\it Dubna, 141 980 Moscow region, Russia}
\vskip5mm
\centerline{${}^{b}$ \it
Laboratoire de Physique Th\'eorique et des Hautes Energies,}
\centerline{\it Unit\'e associ\'ee au CNRS URA 280,~Universit\'e Paris 7}
\centerline{\it 2 Place Jussieu, 75251 Paris Cedex 05, France}
\vskip 1.0truecm  \nopagebreak

\begin{abstract}
\noindent Starting from the most general harmonic superspace
action of self-interacting
$Q^+$ hypermultiplets in the background of $N=2$ conformal supergravity, we
derive the general action for the bosonic sigma model with
a generic $4n$ dimensional
quaternionic-K\"ahler (QK) manifold as the target space.
The action is determined
by the analytic harmonic QK potential and supplies an
efficient systematic procedure of the explicit
construction of QK metrics by the given QK potential.
We find out this action to have two flat limits. One gives the
hyper-K\"ahler (HK) sigma model with a $4n$ dimensional
target manifold, while another yields a conformally-invariant sigma model
with $4(n+1)$ dimensional HK target. We work out the harmonic superspace
version of the QK quotient construction and use it to give a new derivation
of QK extensions of four-dimensional Taub-NUT and Eguchi-Hanson metrics.
We analyze in detail the geometrical and symmetry structure of the second
metric. The QK sigma model approach allows us to reveal the enhancement of
its $SU(2)\otimes U(1)$ isometry to $SU(3)$ or
$SU(1,2)$ at the special relations between its free parameters:
the $Sp(1)$ curvature (``Einstein constant'') and
the ``mass''.

\end{abstract}
\vfill
{\it E-Mail:}\\
{\it 1) eivanov@thsun1.jinr.ru}\\
{\it 2) valent@lpthe.jussieu.fr}

\end{titlepage}

\section{Introduction}
The explicit construction of metrics on the hyper-K\"ahler (HK)
and quaternionic-K\"ahler (QK) manifolds is of interest
both from the purely mathematical point of view
and keeping in mind possible physical
applications of these manifolds in the modern strings and branes stuff.
Important physical multiplets of superbranes
with $N=(1,0)\,,\;d=6$ ($N=2\,,\;d=4$; $N=4\,,\; d=3$; ...)
worldvolume supersymmetry obtained via appropriate compactifications of
M-theory are hypermultiplets (see, e.g. \cite{witt1,ga,ggpt} and
references therein). Their bosonic fields parametrize
HK manifolds in the case of rigid supersymmetry \cite{agf} and
QK manifolds when couplings to the worldvolume supergravity
are turned on \cite{bw}.
As was shown in \cite{ggpt}, the toric HK manifolds \cite{g,ggr}
($4n$-dimensional HK manifolds with $n$ commuting triholomorphic
$U(1)$ isometries) naturally arise in the
$M$-theory context as the $D=11$ supergravity
solutions corresponding to the multiply-intersecting branes.
It would be interesting to construct QK analogues of these
HK manifolds, as well as to reveal their possible branes implications.

A universal method of explicit construction of HK and QK metrics is
the twistor-harmonic one \cite{gios1,gios2,gio}. It emerged
as a natural development of the basic ideas of
the harmonic superspace (HSS) approach to $N=2$ supersymmetric theories
\cite{gikos1,gikos2}. In this method, the defining
constraints of the  HK or QK geometry
are interpreted as integrability conditions for the existence
of some analytic
subspace in an extension of the given $4n$-dimensional
HK or QK manifold by harmonic coordinates on
an internal two-sphere $SU(2)/U(1)$
(in the HK and QK cases they parametrize, respectively, the space
of non-equivalent covariantly constant complex structures and
the $Sp(1)$ component of the holonomy group).
The basic object of HK and QK geometries which fully
specifies the relevant metric (at least, locally)
is the analytic potential,
a function of harmonic variables and other $2n$ coordinates of the
analytic subspace. It is required to have the charge
$+4$ with respect to the harmonic
$U(1)$ (that is the denominator of the harmonic sphere $SU(2)/U(1)$).
Otherwise, it is an arbitrary function
of the involved variables.
As the real advantage of the geometric HSS approach, it provides the practical
recipes of how to construct {\it most general} HK or QK metrics, including
those having no isometries.
Other supersymmetry-inspired approaches to computing such metrics
\cite{hklr}, \cite{gal1} - \cite{gal3} are restricted
to the manifolds with isometries.

In refs.\cite{gios1,gios2} and \cite{gio}
it was shown how to restore the HK or QK metrics by
the known analytic potential. The road from HK or QK potentials
to the metric, as compared, e.g., to that from the K\"ahler potential to
the relevant metric, is somewhat obscured by the necessity to solve differential
equations on the sphere $SU(2)/U(1)$ at the intermediate steps.
In view of lacking of the systematic theory of such equations,
this is the most difficult part of the twistor-harmonic formalism.

In all known examples where the twistor-harmonic approach was
used to compute HK metrics \cite{gios1,gios2,val1,gov},
it was rather easy to solve such equations because the corresponding
HK potentials possessed some triholomorphic isometries.
As for the QK metrics, the relevant equations remain rather complicated
even in the presence of similar isometries (i.e. those which become triholomorphic
in the HK limit, when the $Sp(1)$ curvature vanishes).
In ref. \cite{gio}, only the maximally symmetric case of
homogeneous ${\mbb H}P^n$ manifold was considered as an example.
It corresponds to the vanishing QK potential and is an analog
of the flat HK manifold (and goes into it in the HK limit).
The first example of QK manifold with a non-trivial potential,
a quaternionic extension of the four-dimensional Taub-NUT (TN) manifold,
was studied within the approach of \cite{gio} in \cite{iv1}.
Like in the HK Taub-NUT case \cite{gios1,gios2}, the relevant
harmonic equations can be easily solved due to an $U(1)$ isometry of
the corresponding analytic QK potential.
As the result, the explicit form of the metric
can be readily found.
However, while trying to compute
in this way QK analogs of some other HK metrics allowing
a nice description within the harmonic approach, e.g.
the ``dipolar breaking'' metric \cite{val1,gov}, a serious
technical problem is encountered. It is related
to the necessity to solve one more harmonic differential equation for
the quantity specific just for the QK case.
This is the ``bridge'' connecting harmonic variables
in the original and analytic bases: whereas in the HK case
these variables are the same, it is not so in the QK case.

In view of these subtleties, it is tempting to try some another strategy,
and this is what we do in this paper, taking as the examples QK extensions
of the Taub-NUT (TN) and Eguchi-Hanson (EH) metrics. These QK metrics, in a
rather implicit form, were already constructed in \cite{gal1} proceeding from
the component $N=2$ supergravity (SG)-matter action of ref. \cite{nider}. Their
derivation within the HSS approach not only illustrates this general approach,
but also clarifies some basic features of these metrics, e.g., makes manifest
their symmetry properties.

First of all, we use the lagrangian
version \cite{gios4} of the geometric formalism of ref. \cite{gio}.

Because of the one-to-one correspondence between
the QK manifolds and general matter couplings in $N=2$ SG background \cite{bw},
an action of any such coupling yields a sigma model action
with QK target manifold in the bosonic sector. Conversely,
any QK sigma model action can be lifted to a locally
$N=2$ supersymmetric one. This correspondence becomes manifest
when the $N=2$ sigma model action is formulated
via unconstrained harmonic analytic hypermultiplet
superfields \cite{gios4}. From the standpoint of the
geometric approach of ref. \cite{gio}, these superfields parametrize
the analytic subspace of harmonic extension of the target QK
manifold. The superfield interaction Lagrangian (which in general
is an arbitrary charge $+4$ function of
the hypermultiplet superfields and explicit harmonics) is
just the analytic QK potential. The compensating hypermultiplet
superfield which is needed to descend to
Einstein $N=2$ SG from conformal $N=2$ SG is
the aforementioned bridge relating harmonic variables
in the initial and analytic bases. The bosonic part
of the equations of motion which eliminate an infinite set
of the auxiliary fields coming from the harmonic expansions
of such superfields coincides with the already mentioned
differential equations on $S^2$. Thus this approach can be regarded as
a lagrangian version of the purely geometric approach
of ref. \cite{gio}. As one of its merits,
it directly yields the standard distance
on the target bosonic QK space after elimination
of auxiliary fields, viz. solving the above-mentioned differential
equations on $S^2$. In the geometric approach,
the basic objects are inverse vielbeins and metric.

One of the incentives of this paper is to deduce
a convenient HSS form of the off-shell action of most general QK sigma
model. It was not explicitly given in \cite{gios4}.

Another simplifying device we use here is
the HSS version of the QK quotient construction
(discussed at the component level in \cite{gal1}-\cite{gal3}).
It is a straightforward generalization of the analogous
HSS construction for the HK case \cite{giot} which was used,
in particular, to obtain the EH metric
from a $N=2$ supersymmetric HSS lagrangian\footnote{The $N=1$ superfield version of the HK quotient approach was
earlier employed in \cite{hklr}.}. Basically it amounts to gauging some isometries
of a ``free'' hypermultiplet action by non-propagating
$N=2$ gauge superfields
along the lines of ref. \cite{bgio}. In a manifestly supersymmetric gauge,
after elimination of the gauge superfields by their algebraic
equations of motion,  one ends up with a non-trivial analytic QK potential.
On the other hand, in the Wess-Zumino
gauge, the harmonic differential equations on $S^2$ become linear and can
be immediately solved, thus avoiding the most difficult part of the
geometric twistor-harmonic approach (this concerns as well
the equations for the compensating hypermultiplet in some important cases).
In this gauge, the metric is restored by solving a set
of algebraic equations including some constraints on the original physical
bosonic fields.

This approach is expected to work not only for the
QK generalizations of 4-dimensional EH and TN manifolds,
but also in higher-dimensional cases. So it can hopefully
be used for explicit construction of the QK analogs of metrics on
general toric HK manifolds. We are planning to perform this analysis
in future publications. Our aim here is to explain the basic features of this approach
on the simple 4-dimensional examples: QK extensions of the TN and EH
metrics.

In Sect. 2 we give a short account
of the HSS formulation of the off-shell hypermultiplets actions
in the $N=2$ SG background with focusing
on the bosonic sector which is of primary interest for our purpose of
extracting bosonic QK metrics.
We present a convenient generic
form of the off-shell action (with all fermions discarded) resulting in
the most general target bosonic QK metric.
As the simplest example we firstly derive the action of ${\mbb H}P^n$
sigma model.
Then in Sect. 3 we describe the HSS quotient construction
on the examples of the quaternionic Taub-NUT
and EH metrics. We demonstrate that various isometries of these metrics
have a very transparent form in the HSS lagrangian
language.
We present the final bosonic actions from which
the relevant distances can be read off.
For the TN case the already known result \cite{iv1} is recovered.
In the EH case we still need to solve the appropriate
algebraic constraint. Fortunately, it is the same as in HK case \cite{cf,giot}.
We solve it in Sect. 4 and then study the local and global structure of
the resulting quaternionic EH metric. It is
Einstein with self-dual Weyl tensor and as such has already been studied
in the physical and mathematical literature \cite{hi}-\cite{pp}. The conformal
class encompasses two interesting K\" ahler metrics : the first one is the sclar-flat
Le Brun metric \cite{lb}, while the second one, which seems to be new, has
a non-constant scalar curvature. Both families contain many complete metrics which
are examined. In Conclusions we
summarize the results and outline some problems for future study.

\setcounter{equation}{0}

\section{Quaternionic sigma models from harmonic superspace of local $N=2$ SUSY}

\subsection{The HSS action of $N=2$ SG with matter hypermultiplets}

We start by recalling the salient features of the hypermultiplet
$N=2$ matter coupled to $N=2$ SG in the HSS approach,
basically following refs. \cite{gios4,gio}.
Further details of the HSS approach can be found
in \cite{gikos1,gikos2}.

The $N=2\,, \;D=4$ HSS in the analytic basis
is represented by the following set of coordinates
\beq
\{Z^M\} = \{x^m, \theta^{\hat\mu+}, u^{+i}, u^{-j}, \theta^{\hat\mu-} \}
\equiv \{\zeta^N, \theta^{\hat\mu-}\}~, \;\; \hat\mu = (\mu, \dot\mu),
i,j = 1,2~.
\label{Zzeta}
\eeq
The coordinate set $\{\zeta^N\} = \{x^m, \theta^{\hat\mu+}, u^{\pm i} \}$
forms the analytic harmonic superspace,
harmonic variables $u^{\pm i}$ parametrize a sphere
$S^2$, $u^{+i}u_{i}^- = 1,
\; u^{+i}u^{-k} - u^{+k}u^{-i} = \epsilon^{ki}$
\footnote{We use the following conventions:
$\epsilon_{12} = 1~, \; \epsilon_{ik}\epsilon^{kl} = \delta^l_i$.}.
HSS and its analytic subspace are
real with respect to a generalized conjugation $\widetilde{\;\;}$ which
does not change the harmonic $U(1)$
charges of the HSS coordinates
(it is the product of ordinary complex conjugation and
a Weyl reflection of $S^2$).

The fundamental group of conformal $N=2$ SG
is realized as the analyticity-preserving subclass of diffeomorphisms of
$Z^M$
\beqa
\delta x^m &=& \lambda^m (\zeta)~,\quad \delta \theta^{\hat\mu+} =
\lambda^{\hat\mu+}(\zeta)~, \nn
\delta u^{+ i} &=& \lambda^{++}(\zeta) u^{-i}~, \quad \delta u^{-i} = 0~,
\label{conf1} \\
\delta \theta^{-\hat\mu} &=& \lambda^{-\hat\mu} (\zeta, \theta^-)~.
\label{conf2}
\eeqa

The matter superfields we shall deal with are the analytic
hypermultiplet superfields
$Q^{+r} (\zeta) = F^{ri}(x)u^+_i + ...$, $r = 1,...2n$,
$F^{ri}(x)$ being the physical bosonic fields . These
superfields satisfy the pseudo-reality condition
\beq
\widetilde{(Q^{+}_r)} = \Omega^{rs}\,Q^{+}_s~, \;
\Rightarrow \; \overline{(F^{ri}(x))} =
\Omega_{rs}\epsilon_{ik}\;F^{sk}(x)~, \label{real}
\eeq
where $\Omega_{rs} = -\Omega_{sr}~, \; \Omega^{\,rs}\Omega_{\,st} =
\delta^{\,r}_{\,t}$,
is the totally antisymmetric constant $Sp(n)$ metric. This
condition leaves in $F^{ri}(x)$
just $4n$ real fields which are identified with
the coordinates of target QK manifold. The superfields $Q^{+r}$
are assumed to transform as weight zero scalars under \p{conf1}:
\beq \label{tranQ}
Q^{+r}{}'(\zeta {}') = Q^{+}(\zeta)\;.
\eeq

The full invariant superfield action of the coupled system
of $Q$-hypermultiplets and $N=2$ Einstein
SG consists of the two pieces
\beq
S = S_{SG} + S_{q,Q}~. \label{full}
\eeq

We shall firstly explain the second piece.
It can be written as an integral over the
analytic HSS $\{\zeta^N \}$:
\beq  \label{Sq}
S_{q,Q} = {1\over 2} \int d\zeta^{(-4)}\left\{-q^+_a{\cal D}^{++}q^{+a} +
{\kappa^2\over \gamma^2}
(u^-_i q^{+i})^2 \left[ Q^+_r{\cal D}^{++}Q^{+r} +
L^{+4}(Q^+, v^+, v^-)\right]\right\}~.
\eeq
Here $d\zeta^{(-4)}$ is the measure of integration over $\{\zeta^N \}$
chosen so that $d\zeta^{(-4)}(\theta^+)^4 = d^4x [du]$,
$\kappa $ is the Einstein constant, $\gamma $ ($[\gamma] = -1$) is a sigma
model constant which is set equal to $1$ in what follows,
${\cal D}^{++}$ is the
analyticity-preserving harmonic derivative
covariantized with respect to the group \p{conf1}, \p{conf2}
\beqa
{\cal D}^{++} &=& \partial^{++} + H^{++ m}(\zeta)\partial_m +
H^{++\hat\mu +}(\zeta)\partial_{\hat\mu+} + H^{++\hat\mu -}(\zeta, \theta^-)
\partial_{\hat\mu-} +
H^{+4}(\zeta)\partial^{--} \label{Dcov} \\
\partial^{\pm\pm} &=& u^{\pm i}\frac{\partial}{\partial u^{\mp i}}~,
\quad  \partial_m = \frac{\partial}{\partial x^m}~,
\quad \partial_{\hat\mu \pm} = \frac{\partial}{\partial \theta^{\hat\mu \pm}}~.
\eeqa
The transformation properties of the analytic vielbein
$H^{++m}(\zeta), H^{++\hat\mu+}(\zeta), H^{+4}(\zeta)$
under \p{conf1} are uniquely fixed
by the transformation law of ${\cal D}^{++}$
\beqa
&&\delta {\cal D}^{++} = - \lambda^{++}(\zeta) D^0~,  \label{Dtransf} \\
&& D^0 = \partial^0 +
\theta^{\hat\mu+}\partial_{\hat\mu +}
- \theta^{\hat\mu-}\partial_{\hat\mu -}~, \;\;
\partial^0 = u^{+i}\frac{\partial}{\partial u^{+i}} -
u^{-i}\frac{\partial}{\partial u^{-i}}
\eeqa
(the non-analytic component $H^{++\hat\mu-}$ is pure gauge,
it can be gauged into
$\theta^{\hat\mu+}$ by properly fixing the gauge freedom \p{conf2}).
The irreducible $N=2$ multiplet
carried out by the components of this vielbeins
(modulo the analytic gauge group freedom) is the gauge multiplet
of conformal
$N=2$ SG, the $N=2$ Weyl multiplet.
The object $q^{+ a}(\zeta)$
($a =1,2; \; \widetilde{(q^{+}_a)} = \epsilon^{ab}q^+_b$)
is the compensating hypermultiplet superfield, one of
the two compensating multiplets needed
to descend from $N=2$ conformal SG
to $N=2$ Einstein SG (to the most flexible version of
the latter which allows the general $Q^+$ self-interactions \cite{gios4}).
It transforms so as to cancel the transformation of the analytic
HSS integration measure in \p{Sq}
\beq \label{deltq}
\delta q^{+a}(\zeta) \simeq q^{+a}{}'(\zeta{}') - q^{+a}(\zeta) =
-{1\over 2} \Lambda(\zeta) q^{+a}(\zeta)~, \;\; \Lambda(\zeta) =
\partial_m\lambda^m +
\partial^{--}\lambda^{++} - \partial_{\hat\mu+}\lambda^{\hat\mu+}~.
\eeq
Note the wrong sign of the $q^+$ action compared to that of $Q^+$:
it is the
standard feature of the compensating superfields actions \cite{bibl}.
Also note that in the process of descending to Einstein $N=2$ SG
$q^{+i}$ compensates local conformal $SU(2)$ transformations.

The modified harmonics $v^{\pm i}$ present in the $Q^+$
self-interaction term in
\p{Sq} are related to $u^{\pm i}$ as follows
\beqa
&& v^{+ i} = u^{+ i} - N^{++}(\zeta)u^{-i} = (u^-q^+)^{-1}q^{+i}~, \;\;
N^{++} = {(u^+q^+)\over (u^-q^+)}~, \label{v+} \\
&& v^{-i} = u^{-i}~,
\eeqa
where, from now on,
$$
(u^{\pm}b) \equiv u^{\pm}_ib^i~.
$$
They are invariant under $N=2$ conformal SG group \p{conf1},
\p{deltq} and so can appear inside $L^{+4}$
in any power consistent with the harmonic charge $+4$ of $L^{+4}$.

The flat limit is achieved by putting
\beq
q^{+i} = {1\over |\kappa|}\,u^{+i}\; \;
\Rightarrow \; (u^-q^+) =|\kappa|^{-1}~, \;
N^{++} = 0~, \label{flat}
\eeq
and by equating to zero all the analytic
vielbein components except for
\beq
H^{++m}(\zeta) \;\Rightarrow \;
-2i\theta^{\mu+}\sigma^m_{\mu\dot\mu}\bar\theta^{\dot\mu+}~.
\eeq

It should be pointed out that the action \p{Sq} with an arbitrary $L^{+4}$ yields the
{\it most general} self-interaction of hypermultiplets in the conformal $N=2$ SG
background \cite{gios4}. In the generic case it yields the bosonic sigma model action with the
target QK manifold possessing no any isometries. On the other hand, the component
approach of refs. \cite{nider,gal1,gal2} is limited to the hypermultiplet self-couplings
which give rise to sigma models with certain isometries. This limitation is
related to the use of the conformal compensators with finite sets of
auxiliary fields. Only while using as a compensator the superfield $q^+$ with an infinite set
of auxiliary fields, it becomes possible to construct
the density $\sim (u^-q^+)^2$ which compensates the local supersymmetry transformation
of the analytic superspace integration measure in \p{Sq}. Just this unique property
of $N=2$ SG with $q^+$ as the compensator allows one to arrange an arbitrary
hypermultiplet self-interaction in \p{Sq}.

After this short review of the action $S_{q,Q}$ let us turn
to the pure SG action
$S_{SG}$. It is the action of the compensating
vector multiplet superfield
$H^{++5}(\zeta)$ in the background
of $N=2$ Weyl multiplet (once again, with
the ``wrong'' overall sign):
\beq
S_{SG} = -{1\over 4\kappa^2} \int dZ \;E\,H^{++5}H^{--5}~. \label{Ssg}
\eeq
Here
$dZ~, \; (\;dZ (\theta^+)^4(\theta^-)^4 = d^4x [du]\;)$,
is the measure of integration
over the full HSS. The non-analytic superfield
$H^{--5}(\zeta, \theta^-)$ is defined by
the non-linear harmonic differential
equation
\beq
{\cal D}^{++}H^{--5} - {\cal D}^{--}H^{++5} +{\cal D}^{--}H^{+4}\;H^{--5}
= 0~, \label{H5--}
\eeq
where ${\cal D}^{--}$ is the appropriately covariantized second
harmonic derivative
\beq
{\cal D}^{--} = \partial^{--} + H^{--m}\partial_m +
H^{--\hat\mu+}\partial_{\hat\mu+}
+H^{--\hat\mu-}\partial_{\hat\mu-}~. \label{D--}
\eeq
The non-analytic vielbein components in \p{D--} are determined
by the following equations
\beqa
&& {\cal D}^{++}H^{--m} - {\cal D}^{--}H^{++m} +{\cal D}^{--}H^{+4}\;H^{--m}
= 0~, \label{1--} \\
&& {\cal D}^{++}H^{--\hat\mu\pm} - {\cal D}^{--}H^{++\hat\mu\pm} +
{\cal D}^{--}H^{+4}\;
H^{--\hat\mu\pm} = \pm \theta^{\hat\mu\pm}~. \label{2--}
\eeqa
The non-analytic density $E$ which transforms so as to cancel
the transformations of both the HSS integration measure
and the density $(H^{++5}H^{--5})$ is given by the expression
\beq
E = (-\mbox{det}\,g^{mn})^{-1/2}\;\mbox{det}\,e^{\hat\mu}_{\hat\alpha}
\label{E}
\eeq
with
\beqa
e^{\hat\mu}_{\hat\alpha} &=& \partial_{\hat\alpha-}H^{--\hat\mu+}~,
\label{e} \\
g^{mn} &=& {1\over 32}\;\epsilon^{\hat\alpha\hat\beta\hat\gamma\hat\delta}
e^m_{\hat\alpha\hat\beta}e^n_{\hat\gamma\hat\delta} = {1\over 32}\left(
e^{m \; \beta}_\beta\;e^{n \;\dot\gamma}_{\dot\gamma} -2
e^m_{\alpha\dot\beta}\; e^{n\;\alpha\dot\beta} + (m\leftrightarrow n)\right)~,
\label{g} \\
e^m_{\hat\alpha\hat\beta} &=& \partial_{\hat\alpha-}
\partial_{\hat\beta-}H^{--m} -
\partial_{\hat\alpha-}\,e^{\hat\mu}_{\hat\beta}\,
(e^{-1})^{\hat\nu}_{\hat\mu}\,
\partial_{\hat\nu-}H^{--m}~.
\eeqa

The role of the vector multiplet $H^{++5}$
is to compensate local $\gamma_5$
transformations and dilatations.

\subsection{Generic QK sigma model action}

As was already mentioned, the action $S_{q,Q}$ \p{Sq} with arbitrary
$L^{+4}(Q^+, v^+, v^-)$ is the most general off-shell
action of $n$ self-interacting hypermultiplets
coupled to $N=2$ SG \cite{gios4}. Hence, according
to Bagger and Witten \cite{bw}, in the sector of physical bosons
$F^{ri}(x)$ it should yield the sigma model with a generic $4n$ dimensional
QK manifold as the target space, $F^{ri}(x)$ being local coordinates of
the latter.

Actually, this theorem is manifested by the
above HSS formulation.
This is much like the way how the
one-to-one correspondence between $N=1$ supersymmetric
sigma models and K\"ahler manifolds \cite{zum} is
visualized after writing the off-shell sigma model action
in terms of complex chiral superfields. Similarly,
an analogous correspondence between globally $N=2$
supersymmetric sigma models and HK manifolds \cite{agf} is visualized
when writing the corresponding off-shell sigma model action
via the harmonic analytic $Q^+$ superfields \cite{gikos1,gios2,gio1}.

The isomorphism between locally $N=2$ supersymmetric sigma models and QK
manifolds in the HSS formulation can be seen as follows.
In ref. \cite{gio}, starting from the standard definition of QK
geometry as a constrained Riemannian geometry, it was found that the QK
geometry constraints admit a general solution in terms of unconstrained
potential. This potential is defined on an analytic subspace of harmonic
extension of the initial QK manifold. All the basic geometric objects of
this formulation of QK geometry were proved to be in the one-to-one
correspondence with the quantities appearing in the action \p{Sq}. The
hypermultiplets $Q^{+r}$ are local coordinates of the QK analytic subspace,
the harmonics $v^{\pm i}$ are the relevant harmonic variables, $u^{\pm i}$ are
harmonic variables of the initial harmonic extension of QK manifold (with
$q^{+i}$ being basically the bridge relating these two harmonic sets), the
interaction Lagrangian $L^{+4}$ is just the analytic QK potential which
locally encodes the full information about the related QK metric. This
completes the proof of the one-to-one correspondence between the action
\p{Sq} and the unconstrained description of generic QK manifolds given in
\cite{gio}. Hence \p{Sq} can be used as a tool of explicit construction of
general QK metrics by the given $L^{+4}$.

The problem of extracting QK metric associated
with some $L^{+4}$ can be solved
in an algorithmic way: like in the HK case \cite{gios1,gios2,giot}
it is  reduced to the following two basic steps:
(i) Discarding all fermions; (ii) Passing to the action
of physical bosons $F^{ri}(x)$ by eliminating an infinite
set of auxiliary bosonic fields present in $Q^{+r}$
through their non-dynamical equations of motion.
Important new features as compared to the HK case
are, first, the presence of couplings
to $N=2$ SG in \p{Sq} and, secondly,  the necessity
to take into account the contribution
from the purely SG action \p{Ssg}.

To see in detail how the QK sigma model action arises, let us choose
the WZ gauge for the analytic vielbeins \p{Dcov}
as it was made in \cite{gios4}.
Discarding fermions, in this gauge
\beqa
&& H^{++m}(\zeta) = -2i \,\theta^+\sigma^a\bar\theta^+ \, e^m_a (x)
+ 6(\theta^+)^2(\bar\theta^+)^2\, V^{m\,(ij)}(x)u^-_iu^-_j~, \label{Hm} \\
&& H^{++\mu+}(\zeta) =
(\theta^+)^2 \bar\theta^+_{\dot\mu}\, A^{\mu\dot\mu}(x) +
(\bar\theta^+)^2 \theta^+_{\nu}\, t^{(\nu\mu)}(x)~, \;
H^{++\dot\mu+}(\zeta) =  \widetilde{H^{++\mu+}}(\zeta), \label{Hmu} \\
&& H^{+4}(\zeta) = (\theta^+)^2(\bar\theta^+)^2 \,D(x)
\eeqa
(numerical coefficients are chosen for further convenience).
In Appendix A we show that the fields $A^{\mu\dot\mu}$ (a complex
gauge field for the residual local $\gamma_5$ symmetry and scale transformations)
and $t^{(\nu\mu)}$ do not contribute to the final sigma model action
of the physical bosons $F^{ri}(x)$. So from the beginning
they can be put equal to zero.
Also, we shall not be interested in couplings to gravity, so we put
$$
e^m_a(x) = \delta^m_a
$$
in \p{Hm}. Thus the only components of the Weyl multiplet which couple to
$q^{+i}$ and $Q^{+r}$ in \p{Sq} and can influence the structure of
the final sigma model $F^{ri}(x)$ action
are $D(x)$ and $V^{(ik)}_m(x)$. The former field
is purely auxiliary, the second one
is the gauge field for the local conformal
$SU(2)$ transformations which form a residual symmetry of the WZ gauge
\p{Hm}-\p{Hmu} (this field is also non-propagating).

As the result of this discussion, the part $\tilde{S}_{q,Q}$
of the action \p{Sq} (with $\gamma = 1$) which
is relevant to the calculation of the bosonic QK sigma model
action is obtained from \p{Sq} by discarding all fermionic components in
$q^{+a}(\zeta), Q^{+r}(\zeta)$ and making the substitution
\beq
{\cal D}^{++} \;\Rightarrow \; \tilde{\cal D}^{++} = D^{++} +
(\theta^+)^2(\bar\theta^+)^2\,
\{\; 6\,V^{m\,(ij)}(x)u^-_iu^-_j \,\partial_m + D(x)\,\partial^{--}\; \}~,
\label{substit}
\eeq
where $D^{++}$ is the standard flat harmonic derivative
\beq
D^{++} = \partial^{++} - 2i\theta^+\sigma^m\bar\theta^+\,\partial_m~.
\eeq
Thus
\beq  \label{Sq1}
\tilde{S}_{q,Q} = {1\over 2} \int d\zeta^{(-4)}
\left\{-q^+_a\tilde{\cal D}^{++}q^{+a} + \kappa^2
(u^- q^{+})^2 \left[ Q^+_r\tilde{\cal D}^{++}Q^{+r} +
L^{+4}(Q^+, v^+, v^-)\right]\right\}~.
\eeq
The terms proportional to $D(x)$ and $V^{m\,(ij)}(x)$
in \p{Sq1} can be easily found.
They  have the following universal form
which does not depend on details of $Q^+$ self-couplings
and their coupling to the compensator $q^{+a}$
\beqa
S^{D,V}_{q,Q} &=& {1\over 2}\;\int d^4x \;\{\; D(x)
\int du\; [\;f^{+a}(x,u)\partial^{--}f^{+}_a(x,u) \nn
&& - \;\kappa^2 (u^-f^+)^2F^{+r}(x,u)\partial^{--}F^{+}_r(x,u)
\;] \nn
&& +\; 6V^{m\,(ij)}(x) \int du\, (u^{-}_iu^{-}_j)\;
[\;f^{+a}(x,u)\partial_mf^{+}_a(x,u) \nn
&& - \;\kappa^2(u^-f^+)^2 F^{+r}(x,u)\partial_mF^{+}_r(x,u)
 \;]\; \}\;.
\label{DVq}
\eeqa
Here
\beq
f^{+a}(x,u) \equiv q^{+a}|_{\theta = 0}~, \quad
F^{+r}(x,u) \equiv Q^{+r}|_{\theta = 0}~. \label{defFf}
\eeq

Let us now turn to the supergravity action \p{Ssg}.
It is convenient to choose the WZ gauge
for $H^{++5}(\zeta)$ and to fully fix the
local $\gamma_5$ and dilatation symmetries for which $H^{++5}$ serves
as a compensator, by putting
\beq \label{H5spec}
H^{++5}(\zeta) = i\,\left[(\theta^+)^2 - (\bar\theta^+)^2\right] \;
\Rightarrow \;
H^{--5}(\zeta) = i\,\left[(\theta^-)^2 - (\bar\theta^-)^2\right] + ...~.
\eeq
The bosonic vector gauge field and an $SU(2)$ triplet of auxiliary fields
present in the bosonic part of $H^{++5}$ in the WZ gauge
are of no interest for us because they do not couple to
the matter sector (such couplings could appear only for matter
with non-trivial central charge; this is not the case
for $q^{+i}$ and $Q^{+r}$). The corresponding vielbeins $H^{--5}$
and $H^{--M}$ (entering the harmonic derivative ${\cal D}^{--}$, eq. \p{D--})
are computed using eqs. \p{1--}, \p{2--}, \p{H5--} with the particular
expressions \p{substit}, \p{H5spec} for the analytic harmonic
vielbeins. Then after a straightforward though
tedious computation one finds the only relevant
bosonic term in $S_{SG}$ to be
\beq \label{SGbos}
S_{SG} \; \Rightarrow \;
S_{SG}^{D,V} = -{1\over 2 \kappa^2} \,\int d^4x \left[
D(x) + V^{m(ij)}(x)V_{m(ij)}(x) \right]\;.
\eeq

Now we are in a position to write the generic
HSS action of QK sigma model. It is the sum of \p{Sq1} and \p{SGbos}
\beqa
S_{QK} &=& \tilde{S}_{q,Q} + S_{SG}^{D,V} \nn
&=& {1\over 2} \int d\zeta^{(-4)}\left\{-q^+_a D^{++}q^{+a} + \kappa^2
(u^-_i q^{+i})^2 \left[ Q^+_r D^{++}Q^{+r} +
L^{+4}(Q^+, v^+, v^-)\right]\right\} \nn
&& +\; S_{q,Q}^{D,V} + S_{SG}^{D,V}~.
\label{QKact1}
\eeqa
Varying with respect to the non-propagating
fields $D(x)$ and $V^{m\,(ij)}(x)$
in the last two terms in \p{QKact1} yields, respectively,
the relation between the compensating and matter hypermultiplets
\beq \label{cnstr}
\int du \left[ f^{+a}\partial^{--}f^+_a -\kappa^2(u^-f^+)^2\,
F^{+r}\partial^{--}F^{+}_r \right] ={1\over \kappa^2}
\eeq
and the expression for $V^{m\,(ij)}(x)$
\beqa
&& V_m^{(ij)}(x) = 3\kappa^2 \int du (u^{-i}u^{-j}) V^{++}_m(x,u)
\equiv \kappa^2 {\cal V}_m^{(ij)}(x)~, \label{Vexpr} \\
&& V^{++}_m = f^{+a}\partial_mf^+_a -\kappa^2(u^-f^+)^2\,
F^{+r}\partial_mF^{+}_r~. \label{defV++}
\eeqa
Substituting this back into \p{QKact1} gives the convenient representation
for the QK sigma model action containing only the hypermultiplets fields
\beqa
S_{QK} &=& {1\over 2} \int d\zeta^{(-4)}\left\{-q^+_a D^{++}q^{+a} + \kappa^2
(u^-_i q^{+i})^2 \left[ Q^+_r D^{++}Q^{+r} + L^{+4}(Q^+, v^+, v^-)
\right]\right\} \nn
&& + {\kappa^2\over 2}\int d^4x {\cal V}^{m(ij)}{\cal V}_{m(ij)} \equiv
S_{QK}^{(1)} + S_{QK}^{(2)}~, \label{QKact}
\eeqa
where one should keep only bosonic components
in the superfields $q^{+a}(\zeta)$,
$Q^{+r}(\zeta)$. Also, the constraint \p{cnstr} should be added.

Finally, modulo differences related to the presence
of new terms $\sim \kappa^2$
in the action and the additional constraint \p{cnstr}, the recipe of
deriving the QK metric from \p{QKact} is basically the same as in the
HK case \cite{gios1,val1,gov}. It consists of the following steps.

\vspace{0.3cm}
\noindent{\bf A}. One substitutes into \p{QKact} the $\theta$ expansion of
$q^{+a}$,
$Q^{+r}$ with the fermions discarded
\beqa
q^{+}_a(\zeta) &=& f^{+}_a(x,u) + (\theta^+)^2\,m^{-}_a(x,u) +
(\bar\theta^+)^2\;
\widetilde{m}^{-}_a(x,u) \nn
&& +\; i(\theta^+\sigma^m\bar\theta^+)A^{-}_{ma}(x,u) +
(\theta^+)^2(\bar\theta^+)^2 g^{(-3)}_a(x,u) \nn
Q^{+}_r(\zeta) &=& F^{+}_r(x,u) + (\theta^+)^2\,M^{-}_r(x,u) +
(\bar\theta^+)^2\;
\widetilde{M}^{-}_r(x,u) \nn
&& + \; i(\theta^+\sigma^m\bar\theta^+)B^{-}_{mr}(x,u) +
(\theta^+)^2(\bar\theta^+)^2 G^{(-3)}_r(x,u)~, \label{thetaexp1}
\eeqa
and varies with respect to the unconstrained auxiliary fields
$g^{(-3)}_a$, $G^{(-3)}_r$,
$A^{-}_{ma}$, $B^{-}_{mr}$, $m^{-}_a$, $M^{-}_r$.
Then one inserts the result back
into the action and integrates over the harmonics, thus
expressing the action entirely in terms of the bosonic fields
$f^{ai}(x)$, $F^{ri}(x)$ and their $x$-derivatives. It is easy
to show that the contribution from the fields $m^{-}_a, \widetilde{m}^{-}_a$,
$M^{-}_r, \widetilde{M}^{-}_r$ in the general case
identically vanishes, so from the
very beginning we can put them equal to zero in \p{thetaexp1}
\beq \label{condit}
m^{-}_a = \widetilde{m}^{-}_a = M^{-}_r = \widetilde{M}^{-}_r = 0~.
\eeq

\vspace{0.3cm}
\noindent{\bf B}. One fully fixes the gauge with respect to the
residual local $SU(2)_c$ symmetry of the action \p{QKact} (see below)
so as to gauge away the triplet part of $f^{ai}(x)$:
\beq
f^{i}_a(x) \quad \Rightarrow \quad f^{i}_a(x) =
\delta^i_a\,\sqrt{2}\,\omega(x)~, \label{gauge1}
\eeq
and expresses $\omega(x)$ in terms of the target QK manifold coordinates
$F^{ri}(x)$ using the algebraic constraint \p{cnstr}.

\vspace{0.3cm}
\noindent{\bf C}. One substitutes the expression
for $\omega(x)$ into the previously obtained action
and reads off the QK metric from the latter.

\vspace{0.2cm}
Note that the flat HK limit is achieved by putting altogether
\beq
\sqrt{2}\,|\kappa| \, \omega = 1 \label{HKlim}
\eeq
(recall \p{flat}) and then setting $\kappa = 0$. The constraint \p{cnstr}
becomes identity $1=1$ in this limit, while \p{QKact} goes into the HSS action
of the general off-shell
HK sigma model with the $4n$ dimensional target space \cite{gio1,gios2}:
\beq
S_{QK} \;\Rightarrow \; S_{HK} = {1\over 2} \int d\zeta^{(-4)}\left\{
Q^+_rD^{++}Q^{+r} + L^{+4}(Q^+,u^+,u^-) \right\}~. \label{HKact}
\eeq
We see that, at least at the level of local consideration, any HK manifold
has its QK counterpart and vice-versa.

Before going further, we comment on the peculiarities of realization
of residual local $SU(2)_c$ symmetry in the actions \p{QKact1}, \p{QKact}.
This is the only local symmetry from the SG sector which
survives truncation \p{substit}.
For our further purposes it will be enough to
know its realization on the lowest components $f^{+a}(x,u)$, $F^{+r}(x,u)$
in the expansion \p{thetaexp1}.
Taking into account the general transformation laws \p{tranQ}, \p{deltq}, and
the fact that in the WZ gauge the $SU(2)_c$ gauge parameter
$\lambda^{(ij)}(x)$ appears in the residual coordinate transformations as
\cite{gios4}
\beqa
&& \delta u^{+i} = \lambda^{++}u^{-i} + O(\theta^2)~,\;\; \delta x^m =
O(\theta^2)~,\;\; \delta \theta^{+\hat\mu} = \lambda^{+-}\theta^{+\hat\mu} +
O(\theta^3)~, \label{coorsu2c} \\
&& \lambda^{++} = \lambda^{ij}(x)u^+_iu^+_j~, \;
\lambda^{+-} = \lambda^{ij}(x)u^+_iu^-_j~, \nonumber
\eeqa
we find
\beq
\delta^*\,f^{+a} = \lambda^{+-}\,f^{+a} - \lambda^{++}\,\partial^{--}f^{+a}~,
\quad
\delta^*\,F^{+r} = - \lambda^{++}\,\partial^{--}F^{+r}~. \label{fFsu2c}
\eeq

The invariance of the QK sigma model off-shell action \p{QKact1} under
$SU(2)_c$ is obvious because this action  was
obtained starting from the action \p{full} invariant under the full
$N=2$ SG group \p{conf1}, \p{conf2}, \p{tranQ}, \p{deltq}. In particular, the
invariance of the action $S_{SG}^{D,V}$ follows from
the transformation rules
\beq
\delta V^{ij}_m = -\partial_m\lambda^{ij}(x) + 2\,\lambda^{(i}_{\;\;\;k}(x)\,
V^{j)k}_m(x)~,\;\; \; \delta D(x) = 2 \partial_m
\lambda^{ik}(x)V^m_{ik}(x)~, \label{DVsu2c}
\eeq
which can be easily derived
from the condition of stability of the truncated WZ gauge \p{substit} under
the transformation \p{Dtransf} with $SU(2)_c$ as the only residual local
symmetry.

It is somewhat more tricky to check the invariance
of the pure hypermultiplets action \p{QKact} and the constraint \p{cnstr}.
The latter turns out to be invariant only with taking account of the purely
harmonic (i.e. non-dynamical) equations for $f^{+r}(x,u)$, $F^{+r}(x,u)$.
These equations are obtained by varying with respect to the auxiliary fields
$g^{(-3)}_a$, $G^{(-3)}_r$ defined in \p{thetaexp1} and in the general case read
\beqa
&& \partial^{++}f^{+a} -{\kappa^2\over 2}\,f^{+a}\,\partial_v^{--}L^{+4}
 + \kappa^2\,u^{-a}\,(u^-f^+)\,
 \left(1 - {1\over 2}\,F^{+r}\frac{\partial}{\partial F^{+r}}\right)L^{+4}
= 0~,
 \label{feq} \\
&& \partial^{++}F^{+r}+ F^{+r}\left[{(u^+f^+)\over (u^-f^+)} + {\kappa^2\over
2}\,\partial^{--}_vL^{+4} \right] +
{1\over 2}\,\frac{\partial L^{+4}}{\partial F^{+}_r} = 0~, \label{Feq}
\eeqa
where
$$
\partial^{--}_v = u^{-i}\frac{\partial}{\partial v^{+i}}
$$
acts only on the explicit harmonics $v^{\pm i}$ in $L^{+4}(F^+, v^+, u^-)$.

The result of varying the l.h.s. of the constraint \p{cnstr} with respect
to \p{fFsu2c},
up to a full harmonic derivative, is
\beq
- \int du\;\lambda^{--}\partial^{++} \left[ f^{+a}\partial^{--}f^+_a -
\kappa^2(u^-f^+)^2\,
F^{+r}\partial^{--}F^{+}_r \right]~. \label{varconstr}
\eeq
It is easy to check that eqs. \p{feq}, \p{Feq} imply the ``conservation
law''
\beq
\partial^{++}\left[ f^{+a}\partial^{--}f^+_a -\kappa^2(u^-f^+)^2\,
F^{+r}\partial^{--}F^{+}_r \right] = 0~, \label{conserv1}
\eeq
whence it follows that \p{varconstr} is vanishing, i.e. the constraint
\p{cnstr} is invariant on the shell of non-dynamical $S^2$ equations
\p{feq}, \p{Feq}.
By the way, \p{conserv1} means that the integrand in \p{cnstr} does not
depend on harmonics on-shell and, hence, the harmonic integral
can be taken off in \p{cnstr},
\beq
f^{+a}\partial^{--}f^+_a -\kappa^2(u^-f^+)^2\,F^{+r}\partial^{--}F^{+}_r =
{1\over \kappa^2}~, \label{onshell1}
\eeq
when computing the final form of the action in terms
of the physical bosonic fields $F^{ri}(x)$.

Quite analogously, one can check one more ``conservation law'', this
time for $V^{++}_m(x,u)$ defined in \p{defV++}:
\beq
\partial^{++}V^{++}_m = 0~, \;\Rightarrow \; V^{++}_m(x,u) =
v^{(kl)}_m(x)u^+_ku^+_l~. \label{onshell2}
\eeq
Using this property and eq. \p{onshell1}, it is easy to show that under the
$SU(2)_c$ transformations \p{fFsu2c} the composite gauge field \p{Vexpr} is
transformed just as in \p{DVsu2c}. This is a good check
of self-consistency of our approach. It is also straightforward to check
the $SU(2)_c$ covariance of the system \p{feq}, \p{Feq}.

Finally, we shall make a few further comments on the generality of the above
action. When considering the coupled matter-SG actions within the field
theory framework, the standard assumption is that the kinetic terms of the
physical fields have the correct sign to ensure the positivity of energy.
Otherwise the fields are ghosts. Just to have the standard signs of the
Einstein action and the actions of other physical components of $N=2$ SG and
matter multiplets, the $q$ compensating superfield action in \p{Sq} and that
of $H^{++5}$ \p{Ssg} were chosen to have the ``wrong'' sign as compared to
the sign of the $Q^+$ action in \p{Sq}. However, while using the $N=2$
SG-matter system merely as a tool to recover the QK metrics, the above
requirement is not longer compulsory: we wish to be able to reproduce the
whole set of the QK metrics, with different signatures of the tangent space
metric. In other words, we can choose other relative signs of the $H^{++5}$, $q^+$
and $Q^+$ terms in the whole $N=2$ QK sigma model action: local $N=2$
supersymmetry is always preserved that guarantees the corresponding bosonic
target metric to be QK. The only self-consistency condition we should care
of is related to the constraint \p{cnstr} (or \p{onshell1}). Assuming that
$\kappa^2 = |\kappa|^2 > 0$, eqs. \p{cnstr} or \p{onshell1} in the gauge
\p{gauge1} imply
$$
\omega^2 + \; \ldots = \epsilon\, {1 \over
2\kappa^2}~,
$$
where $\epsilon = \pm 1$ is the relative sign of the $H^{++5}$
and $q^+$ terms and ``dots'' stand for higher orders in $\omega\,(x)$ and $F^{ri}(x)$.
Obviously, for this constraint to be solvable for $\omega(x) = \overline{\omega}(x)$,
one should  demand
\beq
\epsilon = +1~.
\eeq
With this restriction taken into account, eq.
\p{onshell1} implies that
\beq
\omega(x) = {1\over \sqrt{2} |\kappa|}[1 +
\kappa^2\hat{\omega}(\kappa^2, F(x))] \quad \Rightarrow \quad
\partial^m\omega\partial_m\omega = {1\over 2} \kappa^2
\partial^m\hat{\omega}\partial_m\hat{\omega}~,
\label{hatom}
\eeq
where $\hat{\omega}(\kappa^2, F(x))$ is a function of the physical bosonic
fields which is non-singular at $\kappa^2 = 0$.
Then a simple
analysis shows that the above-mentioned uncertainty in the relative signs
amounts to the freedom of changing $\kappa^2 \rightarrow -\kappa^2$
in \p{QKact} (up to the overall sign, and taking into account
that $\kappa^2(u^-q^+)^2
=1 + O(\kappa^2)$), in \p{feq}, \p{Feq}, in the l.h.s. of \p{cnstr}
and \p{onshell1}, and in \p{defV++} . In the ``maximally flat'' case,
with $L^{+4}=0$,
this substitution takes the non-compact
homogeneous hyperbolic target QK manifold
${\mbb H}H^n \sim Sp\,(1,n)/Sp(1)\otimes Sp\,(n)$ into
its compact counterpart ${\mbb H}P^n
\sim Sp\,(1+n)/Sp\,(1)\otimes Sp\,(n)$ (see Subsect.
\ref{example}).

An additional freedom is associated with the possibility to insert
an arbitrary indefinite constant metric in the sum over different
$Q^+_r$ (or $F^+_r$) in all formulas where such a sum is
present. Let us equivalently replace the $Sp(n)$ index
$r$ by the pair $(Ai)$ where $i=1,2$ is a $Sp(1)$ index and $A=1,...n$ is the
vector $SO(n)$ index. Then
$$
Q^+_r{\cal D}^{++}Q^{+r} \sim Q^{+A}_i{\cal D}^{++}Q^{+Ai}
$$
and the generalization just mentioned amounts to the replacement
$$
Q^{+A}_i{\cal D}^{++}Q^{+Ai} \quad \Rightarrow \eta^{AB}Q^{+A}_i{\cal D}^{++}Q^{+Bi}
$$
where $\eta^{AB}$ is the diagonal constant metric with an arbitrary signature.
This corresponds to considering as the ``flat'' backgrounds the homogeneous
QK manifolds $Sp\,(1+m, s)/Sp\,(1)\otimes Sp\,(m,s)$ or
$Sp\,(m, 1+s)/Sp\,(1)\otimes Sp\,(m,s)$, $(m+s=n)$.

In what follows, for definiteness, we shall basically deal
with the QK sigma models
corresponding to the original $N=2$ SG-matter action, i.e. those defined
by the action \p{QKact} and constraint \p{cnstr}, \p{onshell1} with
the fixed signs and $\kappa^2 > 0$. This corresponds to ${\mbb H}H^n$
as the ``maximally flat'' background.

As the last remark, let us notice the existence of another flat limit of the
above general QK sigma model, besides \p{HKact}. It yields the conformally-invariant
HK sigma model with $4(n+1)$ dimensional bosonic target space. One starts from the
action \p{Sq} with the
manifestly included sigma model constant $\gamma$, chooses $\gamma^2 = \kappa^2$
and assumes no flat limit \p{flat} for $q^+_a$. Then putting all $N=2$ SG fields
equal to zero leaves us with the following $N=2$ HK sigma model action
\beqa
S{}'_{HK} &=& {1\over 2} \int d\zeta^{(-4)}\left\{
-q^+_iD^{++}q^{+i} + \tilde{Q}^+_rD^{++}\tilde{Q}^{+r} \right. \nn
&& \left. +\;
(u^-q^+)^2 L^{+4}((u^-q^+)^{-1}\tilde{Q}^+, (u^-q^+)^{-1}q^+, u^-) \right\}~, \label{HKact1}
\eeqa
where
$$
\tilde{Q}^{+r} = (u^-q^+)Q^{+r}~.
$$
and $q^+$ and $Q^+$ are no longer related by any constraint.
One can choose the correct sign for the $q^+$ kinetic term, since $q^+$ ceases to be
the pure gauge compensating superfield after taking this limit. The action
\p{HKact1} (with any sign of the $q^+$ term) is invariant under the rigid $N=2$
superconformal group the realization
of which in the analytic HSS was given in \cite{confA}. It is the most general
superconformally invariant off-shell HSS action of $(n+1)$ self-interacting hypermultiplets.
The on-shell bosonic geometry of such hypermultiplet self-couplings was recently discussed in
\cite{dutch}, \cite{gr2}.

\subsection{Isometries and Killing potential} \label{isometr}
The rigid isometries of HK or QK metrics are nicely represented in the
language of
the HSS actions as symmetries of the latter \cite{bgio}. For further
reference, we briefly describe their realization in the general QK sigma
model HSS action following ref. \cite{bgio}.

Any isometry of the $Q^+$ off-shell action \p{Sq} including those which
become triholomorphic in
the HK limit (in the supersymmetric setting, the latter property
amounts to the commutativity of the
isometry transformations with the rigid $N=2$ supersymmetry) is realized by
the following transformations of $Q^{+r}, q^{+a}$
\beqa
&& \delta Q^{+r} = \epsilon^A \lambda^{+rA}(Q, v)~, \label{Killvect}
\\
&& \delta q^{+a} = \kappa^2 \left[ (u^-q^+)u^{-a}\,
\left(1 - {1\over 2}\,Q^{+r}\frac{\partial}{\partial
Q^{+r}}\right)\Lambda^{++A} -{1\over
2}\,q^{+a}\,\partial_v^{--}\Lambda^{++A} \right] \epsilon^A~,
\label{compenisom} \\
&& \delta v^{+}_i = \kappa^2\, \epsilon^A\,v^-_i
\left(1 - {1\over 2}\,Q^{+r}\frac{\partial}{\partial
Q^{+r}}\right)\Lambda^{++A}~, \label{v+isom}
\eeqa
where
\beq
\lambda^{+rA} = {1\over 2}\,\left(
\frac{\partial \Lambda^{++A}}{\partial Q^{+}_r} +
\kappa^2\,Q^{+r}\,\partial^{--}_v\Lambda^{++A}\right)~, \label{vectpot}
\eeq
and $\Lambda^{++A}(Q,v)$ satisfies the following equation
\beqa
&& \partial^{++}_v\Lambda^{++A} + {1\over 2}\,
\frac{\partial \Lambda^{++A}}{\partial Q^{+r}}\,
\left(\frac{\partial L^{+4}}{\partial Q^{+}_r} +
\kappa^2\,Q^{+r}\,\partial^{--}_v L^{+4}\right) \nn
&& + \kappa^2\,\partial^{--}_v\Lambda^{++A}\,
\left(1 - {1\over 2}\,Q^{+r}\frac{\partial}{\partial
Q^{+r}}\right)L^{+4} - \kappa^2\,\Lambda^{++A}\,\partial^{--}_vL^{+4} = 0~.
\label{Kileq}
\eeqa
The objects $\lambda^{+rA}(Q,v)$ and $\lambda^{++A}(Q,v)$ are called,
respectively, the Killing vector and Killing potential \cite{bgio}.
All possible QK isometries are characterized
by the HSS Killing potential satisfying eq. \p{Kileq}. It is
straightforward to check the invariance of the $Q$ action \p{Sq} under these
transformations.  The $N=2$ SG fields are of course inert under them.
The constraint \p{cnstr} and the composite $SU(2)_c$ gauge filed \p{Vexpr}
can also be easily checked to be invariant under the corresponding
transformations of $f^{+a}$ and $F^{+r}$ (actually without using the $S^2$
equations \p{feq}, \p{Feq} and the Killing potential equation \p{Kileq};
the latter is to be taken into account only when checking the
invariance of the action and covariance of eqs. \p{feq}, \p{Feq}).

\subsection{Example: ${\mbb H}H^n$ and ${\mbb H}P^n$ sigma models}\label{example}
To illustrate the above construction, let us consider the simplest case of the
``flat'' $4n$ dimensional QK manifold, that is
${\mbb H}H^n \sim Sp\,(1,n)/Sp\,(1)\otimes Sp\,(n)$.
It corresponds to the choice
$$
L^{+4} = 0
$$
in \p{QKact} \cite{bgio,gio}. Redefining $Q^{+r}$ as
\beq
Q^{+r} = {1\over \kappa (u^-q+)}\hat{Q}^{+r}~, \label{redef}
\eeq
we cast the superfield lagrangian of $S^{(1)}_{QK}$ in \p{QKact}
into the simple form
\beq \label{hpnQ}
-q^+_aD^{++}q^{+a} + \hat{Q}^+_rD^{++}\hat{Q}^{+r}~.
\eeq
After performing the integration over $\theta^+, \bar\theta^+$
the action $S^{(1)}_{QK}$ is reduced to
\beqa
S^{(1)}_{QK} &=& \int d^4xdu \left( \partial^{++} f^+_a g^{(-3)a} -
A^-_{ma}\partial^mf^{+a} + {1\over 4} A^-_{ma}\partial^{++}A^{-ma}
\right.\nn
&& \left. -\;\partial^{++} \hat{F}^+_r \hat{G}^{(-3)r} + \hat{B}^-_{mr}
\partial^m\hat{F}^{+r} - {1\over 4} \hat{B}^-_{mr}\partial^{++}\hat{B}^{-mr}
\right)~. \label{HP1}
\eeqa
Varying with respect to the auxiliary fields in this case yields
\beqa
&& \delta g^{(-3)}_a: \qquad \partial^{++} f^{+a}(x,u) = 0 \;\Rightarrow \;
f^{+a}(x,u) = f^{ai}(x)u^+_i = \sqrt{2}\, \epsilon^{ai}\,\omega(x)\,u^+_i~,
\label{fexpr} \\
&& \delta A^{-}_{ma}: \qquad \partial^{++}A^{-a}_m -2\,\partial_mf^{+a} =
0 \;\Rightarrow \;
A^{-a}_{m} = 2\,\sqrt{2}\,u^{-a}\, \partial_m \,\omega(x)~,
\label{Aexpr} \\
&&\delta \hat{G}^{(-3)}_r: \qquad \partial^{++} \hat{F}^{+r}(x,u) = 0 \;
\Rightarrow \; \hat{F}^{+r}(x,u) = \hat{F}^{ri}(x)u^+_i~, \label{Fexpr} \\
&& \delta \hat{B}^{-}_{mr}: \qquad
\partial^{++}\hat{B}^{-r}_m -2\,\partial_m\,\hat{F}^{+r} = 0 \;\Rightarrow \;
\hat{B}^{-r}_m =
2\partial_m \hat{F}^{ri}(x)u^-_i~. \label{HP2}
\eeqa
After substituting this back into the action and integrating over harmonics,
one gets
\beq
S^{(1)}_{QK} =
\int d^4 x \;\{\; {1\over 2}\partial_m\hat{F}_{ri}\partial^m\hat{F}^{ri}
-2 \partial_m\omega(x) \partial^m\omega(x)\; \}~. \label{HP3}
\eeq
The relevant expressions for $\omega(x)$ and ${\cal V}_{m}^{(ij)}$
following from eqs. \p{cnstr}, \p{Vexpr}, \p{defV++}  are also easy to find
\beq
\omega(x) = {1\over \sqrt{2}\kappa}
\sqrt{1 + {\kappa^2\over 2}\,\hat{F}^2}~, \quad
{\cal V}_{m}^{(ij)} = \hat{F}_{r}^{(i}\partial_m \hat{F}^{rj)}~. \label{HP5}
\eeq
Finally, the action \p{QKact} for this simple case is given by
\beqa
S_{HP} &=& {1\over 2} \int d^4x \;\{\; (\partial_m\hat{F}\partial^m\hat{F}) +
\kappa^2\;(\hat{F}_{r(i}\partial_m \hat{F}^r_{j)})(\hat{F}_{s}^{(i}\partial^m
\hat{F}^{sj)}) \nn
&& - \;{\kappa^2\over 2}\,{1\over 1 + {\kappa^2\over 2}\, \hat{F}^2}
(\hat{F}\partial_m\hat{F})(\hat{F}\partial^m\hat{F}) \;\}~, \label{HP6}
\eeqa
or, in terms of the original variables $F^{ri} =
(\sqrt{2} \kappa \omega)^{-1} \;\hat{F}^{ri}$,
\beq
S_{HP} = {1\over 2} \int d^4x
\;\{\; {1\over 1-{\kappa^2\over 2}\,F^2}(\partial_m F\partial^m F) +
{\kappa^2\over [1 - {\kappa^2\over 2}\, F^2]^2}
(F_{ri}F^i_s)(\partial_m F^r_j \partial^m F^{sj})\;\}~. \label{HP7}
\eeq
The corresponding metric is recognized as that of the homogeneous
non-compact space
$\dst{\mbb H}H^n \sim Sp\,(1,n)/ Sp\,(1)\otimes Sp\,(n)$, or of
its compact counterpart
${\mbb H}P^n \sim Sp\,(1+n)/Sp\,(1)\otimes Sp\,(n)$ for
$\kappa^2 \rightarrow - \kappa^2$.

The $Sp\,(1,n)$ isometry of the metric is originally realized as the following
transformations of the superfields $q^{+a}$ and $Q^{+a}$ \cite{bgio}
\beqa
Sp\,(1):  && \delta Q^{+r} = \kappa^2\,
\beta^{(ij)}\,(v^+_iv^-_j)\,Q^{+r}\,, \; \delta q^+_i = \kappa^2\,
\beta_{(ik)}\,q^{+k}\,, \label{sp1} \\
Sp\,(n):  && \delta Q^{+r} = \lambda^{(rs)}Q^+_s\,, \quad \delta q^{+a} =
0~, \label{spn} \\
Sp\,(1,n)/Sp\,(1)\otimes Sp\,(n):  && \delta Q^{+r} = \lambda^{ri}v^{+}_i +
\kappa^2\,\lambda^{si}\,(Q^+_sv^-_i)\,Q^{+r}\,, \nn
&& \delta q^{+i} = \kappa^2\,\lambda^{si}\,Q^+_s\,(v^-q^+)~. \label{spcoset}
\eeqa
The corresponding Killing potentials are
\beq
\Lambda^{++(ik)} = v^{+i}v^{+k}~, \quad \Lambda^{++(rs)} = Q^{+r}Q^{+s}~,
\quad \Lambda^{++ri} = 2\,v^{+i}Q^{+r}~. \label{sppoten}
\eeq
They are evident solution of the Killing equation \p{Kileq} which
is simply
$$
\partial_v^{++}\Lambda^{++A} = 0
$$
since $L^{+4} = 0$ in this case. The $Sp\,(1,n)$
transformation properties of $f^{+a}$, $F^{+r}$
and, further, of the physical fields $F^{ri}(x)$ can be easily derived from
\p{sp1} - \p{spcoset} using \p{fexpr}, \p{Fexpr} and taking into account the
compensating $SU(2)_c$ transformation needed to preserve the
$SU(2)_c$ gauge \p{gauge1}.
\setcounter{equation}{0}

\section{QK quotient construction in the HSS approach}
\subsection{HSS quotient for the quaternionic Taub-NUT}
Having the general $N=2$ SG inspired representation \p{QKact}, \p{cnstr}
for the QK sigma model action, we are guaranteed, by the reasoning of
\cite{gio}, that the physical bosonic fields target metric is QK
for any choice of the analytic QK potential $L^{+4}(F, v)$, as soon as
the infinite set of bosonic auxiliary fields has been eliminated by the
non-dynamical harmonic equations \p{feq}, \p{Feq} and the $SU(2)_c$
gauge \p{gauge1} has been imposed. Thus the action \p{QKact} augmented
with the algebraic constraint \p{cnstr} (or its on-shell form \p{onshell1})
provides us with a {\it general} tool of computing QK metrics by the given $L^{+4}$,
in full analogy to the similar HSS approach to computing HK metrics
\cite{gios1,gios2,val1,gov}.
However, as was already mentioned earlier, for the generic $L^{+4}$ eqs. \p{feq},
\p{Feq} are difficult to solve. The situation in the QK case is more
complicated than in the HK one due to the presence of the extra harmonic equation \p{feq} for
the compensating hypermultiplet; in general, \p{feq} and \p{Feq} form a
highly nonlinear coupled system of differential equations on the sphere
$S^2 = \{u^{\pm i}\}$.

Like in the HK case, in a number of interesting examples there is a way round
this difficulty. This is the HSS version of the
quotient construction \cite{hklr,gal1,gal2}.
It is based on the observation that
many non-trivial HSS actions with the isometries of the type
discussed in Subsect.\ref{isometr} (both in the HK and QK cases)
can be thought of as gauge-fixed forms of
more general gauge invariant actions including non-propagating HSS vector
gauge multiplets. The simplest non-trivial example of this sort is provided
by the QK extension of the four-dimensional Taub-NUT metric, and
we shall illustrate the method just on this example.

The HSS action for this case was written for the first time in \cite{bgio}.
It is defined by the same quartic analytic potential $L^{+4}$ as in the HK
case \cite{gios1}
\beq \label{TNpot}
L^{+4}_{TN} = {1\over 4}\,(c^{ab}Q^+_aQ^+_b)^2 \equiv
{1\over 4}\,(J^{++})^2~, \;\; a,b = 1,2~.
\eeq
Here $c^{(ab)}$ is a constant vector
which breaks the Pauli-Gursey $SU(2)_{PG} \sim Sp(1)_{PG}$ isometry of the free $Q^{+a}$
action (with the Killing potential $Q^{+a}Q^{+b}$) down to $U(1)_{PG}$ (with the
Killing potential $c^{ab}Q^+_aQ^+_b$). The coefficient $1/4$
was chosen for further convenience. The full symmetry of the QK Taub-NUT
HSS action is $Sp(1)\times U(1)_{PG}$, with $Sp(1)$ given by \p{sp1}.
Assuming $c^{(ab)}$ to obey the standard pseudo-reality condition
\beq
\overline{c^{(ab)}} = \epsilon_{ad}\epsilon_{be}c^{(de)}~,
\eeq
which together with \p{real} implies $J^{++}$ to be real, $\widetilde{J^{++}} =
J^{++}$, and choosing the $SU(2)_{PG}$ frame so that
\beq
c^{11} = c^{22} = 0~, \quad c^{12} = i \lambda~,
\quad \overline{\lambda} = \lambda~,
\label{frame0}
\eeq
one can rewrite \p{TNpot} as
\beq
L^{+4}_{TN} = -\lambda^2\,(Q^+ \bar Q^+)^2~, \quad Q^{+} \equiv Q^+_1~,
\;\; \bar Q^{+} = - Q^+_2~. \label{TNpot1}
\eeq
This form of the QTN potential was used in \cite{bgio,iv1}.

The QTN metric in the harmonic approach was explicitly found in
\cite{iv1} by making use of the pure geometric formalism of \cite{gio}.
It is easy to show that, up to a slight difference in the notation,
the general $S^2$ equations \p{feq}, \p{Feq} in this case
coincide with the equations for the harmonic and $x^a, \bar x^a$- bridges
employed in ref. \cite{iv1}.  Therefore the lagrangian approach we
prefer to
deal with here yields the same answer for the metric as the approach used
in \cite{iv1}.
We are not going to compare both approaches in detail, we merely wish
to show
that the lagrangian approach suggests an alternative method to derive
the QTN metric.

Let us start from the system of two free hypermultiplets $Q^{+a}, G^{+b}$,
namely from
the superfield Lagrangian which corresponds to a 8-dimensional
${\mbb H}H^2$ manifold (cf. \p{hpnQ})
\beq \label{twoTN}
{\cal L}^{+4} = -q^+_a{\cal D}^{++}q^{+a} +\kappa^2
(u^-q^+)^2\left[ Q^+_a{\cal D}^{++}Q^{+a} + G^{+}_a{\cal D}^{++}G^{+a}
\right]~.
\eeq
Now, let us specialize to the following one-parameter isometry
of this action
\beqa
&& \delta Q^{+a} = \epsilon \left[ c^{(ab)}Q^{+}_b -
\kappa^2\,(u^-G^+)\,Q^{+a}
\right]~, \nn
&& \delta G^{+a} = \epsilon \left[ v^{+a} -
\kappa^2\,(u^-G^+)\,G^{+a}\right]~, \nn
&& \delta q^{+i} = \epsilon\,\kappa^2\left[ (u^-G^+)\,q^{+i} -
(v^+G^+)\,(u^-q^+)\,u^{-i} \right]~. \label{isom1}
\eeqa
The Killing potential for this particular case of the ${\mbb H}H^2$ isometries
(see \p{Killvect} - \p{Kileq}) can be easily found
\beq
\Lambda^{++} = 2\,(v^+G^+) - c^{(ab)}\,Q^+_aQ^+_b = 2\,(v^+G^+) - J^{++}~.
\label{Kp1}
\eeq
Let us now {\it gauge} this isometry by promoting $\epsilon$ to an analytic
gauge parameter, $\epsilon \;\Rightarrow \; \epsilon\,(\zeta)$.
As was shown in
\cite{bgio}, the gauge invariant lagrangian is the following
modification of \p{twoTN}
\beq
{\cal L}^{+4}{}' =
-q^+_a{\cal D}^{++}q^{+a} +\kappa^2
(u^-q^+)^2\left[ Q^+_a{\cal D}^{++}Q^{+a} + G^{+}_a{\cal D}^{++}G^{+a} +
V^{++}\Lambda^{++} \right]~, \label{gaugeL}
\eeq
where $V^{++} = V^{++}(\zeta)$ is the standard analytic $N=2$ gauge
potential with the transformation law
\beq \label{isomV}
\delta V^{++} = {\cal D}^{++}\epsilon\,(\zeta) ~.
\eeq
It is straightforward to check invariance of the modified $Q, G$ action
under \p{isom1}, \p{isomV} with local $\epsilon\,(\zeta)$

The $N=2$ vector gauge multiplet is assumed to be non-propagating, i.e.
having no kinetic term. In other words, it is the Lagrange
multiplier for the constraint
\beq \label{constrTN}
\Lambda^{++} =
2\,(v^+G^+) - J^{++} = 0~.
\eeq
Further, the transformation law \p{isom1}
shows that one can fully fix the $\epsilon(\zeta)$ gauge freedom by choosing
the gauge
\beq (u^-G^+) = 0 \;\Rightarrow \; G^{+a} = -(v^+G^+)\,u^{-a}~. \label{gauge2}
\eeq
In this gauge
\beq
G^+_a{\cal D}^{++}G^{+a} = (v^+G^+)^2~. \label{kingau2}
\eeq
Expressing $(v^+G^+)$ from the constraint \p{constrTN},
$$
(v^+G^+) = {1\over 2}J^{++}~,
$$
and substituting this back into \p{gaugeL} with
taking account of \p{kingau2}, we recover the QTN potential \p{TNpot}. Note
that we could choose the sign minus in front of the $G$ part of the action \p{twoTN}
and properly modify the isometry group (it is related to \p{isom1}
via the substitution $\kappa^2
\rightarrow -\kappa^2$). This would give rise to QTN potential with the sign
minus as compared to \p{TNpot}, \p{TNpot1}. These two choices are
related by the change $\lambda^2 \rightarrow -\lambda^2$ and yield
the metrics with the positive and negative Taub-NUT ``mass'' which is basically
just $\lambda^2$.

So we started from the system of two ``free'' $Q$ hypermultiplets, gauged
its some $U(1)$ isometry by the non-propagating gauge potential $V^{++}$
and succeeded in eliminating one of these hypermultiplets by choosing
the appropriate gauge and solving the constraint implied by $V^{++}$ as the
Lagrange multiplier. As the result we have gained the QTN action for the
remaining hypermultiplet. In other words, the QTN action proves to be
a special gauge of the more general action \p{gaugeL}.

We are at freedom to choose other gauges in \p{gaugeL}, the
final bosonic sigma model action should evidently be the same
irrespective of the gauge-fixing procedure. A
convenient gauge is the Wess-Zumino (WZ) gauge for $V^{++}$. Discarding fermions
and the terms $\sim (\theta^+)^2, (\bar\theta^+)^2$ (their contribution
finally disappears from the action), $V^{++}$ in this gauge reads
\beq
V^{++}_{WZ}(\zeta) = i(\theta^+\sigma^n\bar\theta^+)\,A_n(x) +
(\theta^+)^2(\bar\theta^+)^2\,S^{(ik)}(x)u^-_iu^-_k~. \label{WZ1}
\eeq
The residual gauge freedom is given by the transformations
\beq \label{resid}
\delta A_m(x) = -2\,\partial_m \,\epsilon\, (x)~, \qquad \epsilon\, (\zeta) =
\epsilon \,(x) + ...~.
\eeq

Let us pass in \p{gaugeL} to the superfields $\hat Q^+$, $\hat G^+$
defined in eq. \p{redef}:
\beq \label{hatTN}
{\cal L}^{+4}_{gauge} =
-q^+_a{\cal D}^{++}q^{+a} + \hat{Q}^+_a{\cal D}^{++}\hat{Q}^{+a} +
\hat{G}^{+}_a{\cal D}^{++}\hat{G}^{+a} +
V^{++}_{WZ}\left[\, 2\,(q^+\hat{G}^+)\, - \hat{J}^{++}\,\right]~.
\eeq
An important feature of WZ  gauge is that the hypermultiplet
auxiliary fields $\hat{g}^{(-3)}_i (x,u)$ and $ \hat{G}^{(-3)}_r(x,u)$
(see \p{thetaexp1}) drop out from the interaction term in \p{hatTN}
in view of the structure of
$V^{++}_{WZ}$ \p{WZ1}. As the result, the $S^2$ equations obtained by
varying with respect to these fields are {\it linear} in this gauge
\beqa
&& \partial^{++}f^+_a = 0\;, \;\Rightarrow \; f^{+}_a =
\sqrt{2}\,\delta^i_a\,\omega(x)
u^+_i~, \nn
&& \partial^{++}\hat{F}^{+a} = \partial^{++}\hat{g}^{+a} = 0\,,\;
\Rightarrow \; \hat{F}^{+a} = \hat{F}^{ia}(x)u^+_i\;, \;\;
\hat{g}^{+a} = \hat{g}^{ia}(x)u^+_i~, \label{S2TN}
\eeqa
where $\hat{g}^{+a} = \hat{G}^{+a}|_{\theta = 0}$. After substituting
this solution back into
\p{QKact1} (or \p{QKact} and \p{cnstr}) the problem of finding the action of physical
bosons is reduced to eliminating the remainder of auxiliary fields,
fixing the residual gauge freedom
\p{resid} by the gauge condition
\beq
\hat{g}^{ia}(x) = \hat{g}^{(ia)}(x)~, \qquad (\mbox{or} \quad
\epsilon_{ia}\,\hat{g}^{ia} = 0)
\eeq
and employing the algebraic constraint
\beq
\hat{g}^{(ik)}(x) = {1\over 2\,\sqrt{2}\,|\kappa| \omega(x)}\,
c^{(ab)}\hat{F}^i_a(x)\hat{F}^k_b(x) \quad \mbox{or} \quad
g^{(ik)}(x) = {1\over 2}\,
c^{(ab)}F^i_a(x)F^k_b(x)~,
\label{constrg}
\eeq
which is obtained  by varying with respect to the Lagrange multiplier $S^{(ik)}\,(x)$ in the
component form of the action corresponding to \p{hatTN}.
The fields $\omega\,(x)$ and ${\cal V}^{(ik)}_m\,(x)$ are
expressed by the general relations \p{cnstr}, \p{Vexpr}:
\beq
\omega = {1\over \sqrt{2}|\kappa|}\sqrt{1 +{\kappa^2\over
2}\,\hat{g}^2}~, \qquad
{\cal V}^{(ij)}_m = \hat{g}_{a}^{(i}\,\partial_m\hat{g}^{j)a}~.
\label{omVqtn}
\eeq

The $S^{(1)}_{QK}$ part of the general action \p{QKact} in the present case,
\beq
S^{(1)}_{QTN} = {1\over 2} \int d\zeta^{(-4)} {\cal L}^{+4}_{gauge}~,
\eeq
after integration over $\theta$s and eliminating auxiliary fields in $q^+, Q^+,
G^+$ becomes
\beqa
&& S^{(1)}_{QTN} = \int d^4x \left( {\cal L}^{kin} + {\cal L}^{vect}
\right)~,
\nn
&& {\cal L}^{kin} = {1\over 2}\partial^m \hat{F}^{ia}\partial_m \hat{F}_{ia} +
 {1\over 2}\partial^m \hat{g}^{ik}\partial_m \hat{g}_{ik}  -
 2 \partial^m \omega\partial_m \omega~, \label{kin1} \\
&& {\cal L}^{vect} = {1\over 2} \left\{ V^m \,c^{(ab)}\hat{F}^i_a\partial_m
\hat{F}_{ib} + V^2 \left[\kappa^2\omega^2 +{1\over 8}\,c^2\, \hat{F}^2
-{\kappa^2\over 4}\,\hat{g}^2\right]\right\}~, \label{vect1}
\eeqa
where
$$
c^2 \equiv c^{(ab)}c_{(ab)} = 2 \lambda^2~.
$$

Recall that the full QTN sigma model action is given by (see \p{QKact})
\beq
S_{QTN} = S^{(1)}_{QTN} + S^{(2)}_{QTN} =
\int d^4x \left( {\cal L}^{kin} + {\cal L}^{vect} + {\kappa^2\over 2} {\cal V}^2
\right)~. \label{QTNact}
\eeq
In order to get the final physical bosons action,
one should pass to
$$
g^{(ik)} = {1\over \sqrt{2} |\kappa|\omega}\hat{g}^{(ik)}~, \qquad
F^{ai} = {1\over \sqrt{2} |\kappa|\omega}\hat{F}^{ai}~,
$$
define
\beqa
&& F^{1i} \equiv \phi^i~, \quad \bar{\phi}_i = \overline{F^{1i}} = F^2_i = F_{1i}~,
\quad s = \bar s \equiv \phi^i\bar\phi_i \;\Rightarrow \nn
&& g^{(ik)} = -i\lambda\,\phi^{(i}\bar\phi^{k)}~, \quad
F^2 = 2 s~, \quad g^2 = {\lambda^2\over 2} \,s^2~, \label{defphi}
\eeqa
substitute all this (together with \p{omVqtn}) into \p{QTNact}, and finally
eliminate the non-propaga\-ting field $V_m\,(x)$ by its equation of
motion. After these purely algebraic manipulations the final form of the
action is as follows
\beqa
S_{QTN} &=& {1\over 2} \int d^4 x \left(
g_{1i\,1k}\partial^m\phi^i\partial_m\phi^k +
g_{2i\,2k}\partial^m\bar\phi^i\partial_m\bar\phi^k \right. \nn
&& \left. + \;
g_{1i\,2k}\partial^m\phi^i\partial_m\bar\phi^k +
g_{2i\,1k}\partial^m\bar\phi^i\partial_m\phi^k \right)~,
\eeqa
with
\beqa
&& g_{1i\,1k} = {A\over B\, C^2}\; \bar\phi_i\bar\phi_k~, \quad
g_{2i\,2k} = {A\over B\, C^2}\; \phi_i \phi_k~, \nn
&& g_{1i\,2k} = g_{2k\,1i} ={1\over B\, C^2}
\left(\epsilon_{ik}\; B^2 + A\;\bar\phi_i\phi_k \right)~, \nn
&& A = {\lambda^2}\,(1-{\kappa^2\over 2}\,s)\,(1 +{\lambda^2\over 2}\,s)~,
\quad
B = 1 +\lambda^2\,s - {\kappa^2\over 4}\,\lambda^2\,s^2~, \nn
&& C = 1 -\kappa^2 \left(s + {\lambda^2\over 4}\,s^2\right)~.
\label{QTNmetric}
\eeqa

This expression for the QTN metric coincides with the one obtained
in \cite{iv1} (up to the sign of $\kappa^2$ and a normalization of
$\lambda^2$) and with the metric following from the HSS sigma model
action with $L^{+4}$ defined by eqs. \p{TNpot}, \p{TNpot1}
\footnote{Presumably, it is also related, via a change of parametrization,
to the QK extension of TN metric derived in \cite{gal1} within the component version of the
QK quotient construction.}. The computation using the HSS quotient
approach turns out much simpler as it allows to avoid the problem of solving
nonlinear differential equations on the harmonic sphere $S^2$.

Finally, we note that the ``master'' gauge-invariant action corresponding to
the Lagrangian \p{gaugeL} possesses global $SU(2)\otimes U(1)$ symmetry with
the Killing potentials
\beq
\Lambda^{++}_{ik} = v^+_iv^+_k - \kappa^2\,G^+_iG^+_k~, \qquad \Lambda^{++}
= J^{++} \label{QTNisom}
\eeq
(these combinations of the free action symmetries are singled
out by the condition of invariance of the interaction term). This symmetry
just amounts to the familiar $SU(2)\times U(1)$ isometry of the QTN metric.

\subsection{Quaternionic EH metric from the HSS quotient}
After explaining the basic points of the HSS quotient construction on the TN
example let us discuss another interesting example of QK metric which was
not still explicitly worked out in the HSS approach. It is QK generalization
of another famous four-dimensional HK metric, the Eguchi-Hanson (EH) one
\cite{egh,eh}. Actually, this HK metric was explicitly constructed in the
HSS approach starting just from the appropriate quotient construction
\cite{giot}. A QK generalization of the corresponding gauged $Q^+$ action
was given in \cite{bgio}.

Like in the QTN example, our starting point is the
${\mbb H}H^2$ action of two hypermultiplets
$Q^{+aA}(\zeta)$, $A=1,2$, but the
abelian isometry to be gauged will be $SO(2)$ which rotates
$Q^{+aA}$ through each
other with respect to the index $A$. To be more precise, the QK
generalization of the ``master'' EH Lagrangian of ref. \cite{giot} reads
\cite{bgio}
\beq {\cal L}^{+4}_{EH} = - q^+_i{\cal D}^{++}q^{+i} +
\kappa^2\,(u^-q^+)^2\left[\, Q^{+B}_a{\cal D}^{++}Q^{+aB} +
V^{++}\left(\epsilon^{AB}Q^{+aA}Q^{+B}_a + \xi^{++} \right)\right]~,
\label{EHmast}
\eeq
where
\beq \label{real2}
\xi^{++} = \xi^{(ik)}v^+_iv^+_k~, \qquad \overline{(\xi^{(ik)})} =
\epsilon_{il}\,\epsilon_{kj}\,\xi^{(lj)}~.
\eeq
It is
invariant under the following $SO(2)$ gauge transformations with the local
parameter $\varepsilon(\zeta)$:
\beqa
&& \delta Q^{+A}_a = \varepsilon
\,\left[\, \epsilon^{AB}Q^{+B}_a -
\kappa^2\,\xi^{+-}\,Q^{+A}_a \,\right]~,\nn
&& \delta q^{+i} = \varepsilon\,\kappa^2 \left[\,\xi^{+-}\,q^{+i}
- \xi^{++}
(u^-q^+)\,u^{-i}\,\right]~, \label{EHisom} \\
&& \delta V^{++} = {\cal D}^{++}\,\varepsilon~.
\eeqa
It is straightforward to see that the
supercurrent to which $V^{++}$ couples in \p{EHmast} is just the Killing
potential for the rigid subgroup of these transformations.

The passing to the physical bosons component action follows the steps {\bf
A} - {\bf C} of Subsect. 2.2, with
\beq
{\cal L}^{+4\;(1)}_{EH} =
- q^+_i D^{++}q^{+i} +
\hat{Q}^{+B}_a D^{++} \hat{Q}^{+aB} +
V^{++}_{WZ}\left(\epsilon^{AB}\hat{Q}^{+aA}\hat{Q}^{+B}_a
+ \kappa^2\,\xi^{(ik)}q^+_iq^+_k \right)~. \label{EH1}
\eeq
We have fixed the WZ gauge \p{WZ1} for $V^{++}$ and redefined the
hypermultiplet superfields according to \p{redef}.

Now one should substitute  the purely
bosonic $\theta$ expansions \p{thetaexp1}, \p{WZ1}
for the involved superfields (with \p{condit} taken into account).
After integrating over $\theta$s,  the free part of this action precisely
coincides with \p{HP1}, so after varying with respect to $\hat{g}^{(-3)a}$,
$\hat{G}^{(-3)aA}$
(these auxiliary fields do not appear in the interaction part) one gets for
$\hat{f}^{+a}, \hat{F}^{+aA}$ the linear harmonic equations like in the QTN
case
\beq
\partial^{++}f^{+a} = 0 \;\Rightarrow \; f^{+a} =
\sqrt{2}\,\epsilon^{ai}\,\omega(x)u^+_i\, \quad
\partial^{++}\hat{F}^{+aA} = 0 \;\Rightarrow \; \hat{F}^{+aA} =
\hat{F}^{aiA}(x)u^+_i\;. \label{EHs2}
\eeq
The full action $S^{(1)}_{EH}$ then reads
\beqa
S^{(1)}_{EH} = &=& \int d^4xdu \left\{-
A^-_{ma}\partial^mf^{+a} + {1\over 4} A^-_{ma}\partial^{++}A^{-ma}
+ \hat{B}^{-A}_{ma}
\partial^m\hat{F}^{+aA} \right. \nn
&& \left.  - \; {1\over 4} \hat{B}^{-A}_{ma}\partial^{++}\hat{B}^{-maA} +
{1\over 2}\, S^{(ik)}u^-_iu^-_k\left[\hat{F}^{+aB}\hat{F}^{+A}_a\epsilon^{BA}
+ 2\,\kappa^2 \omega^2\,\xi^{++} \right] \right. \nn
&& \left.   - \;{1\over
2}\,A^m\left[ \hat{F}^{+bA}\hat{B}^{-C}_{mb}\epsilon^{AC} + \sqrt{2}\,\kappa^2\omega
\left( (u^-A^-_m)\,\xi^{++} - (u^+A^-_m)\,\xi^{+-} \right)\right]\right\}
\label{EH10}
\eeqa
where now $\xi^{++}=\xi^{ik}u^+_iu^+_k,\;
\xi^{+-}=\xi^{ik}u^+_iu^-_k$. Varying it with respect to the fields $A^-_{ma}(x,u)$,
$\hat{B}^{-A}_{ma}(x,u)$ allows one to expresses them in terms of the remaining fields
\beqa
&& A^{-a}_m = 2\,\sqrt{2}\,u^{-a}\,\partial_m \,\omega + \sqrt{2}\,\kappa^2\omega \,
\xi^{(ai)}u^-_i\,A_m \nn
&& \hat{B}^{-aA}_m =
2\,\partial_m\,\hat{F}^{aiA}\,u^-_i - A_m\,\hat{F}^{aiB}u^-_i\,
\epsilon^{BA}~,
\label{vect}
\eeqa
while varying with respect to $S^{ik}(x)$ produces the
algebraic constraint
\beq
\hat{F}^{Aa(i}\hat{F}^{Bj)}_a +2\,\kappa^2\omega^2\,\xi^{(ij)} = 0\,.
\label{EHcnstr}
\eeq
Substituting these expressions back into \p{EH10}, we finally get
\beqa
S^{(1)}_{EH} = \int d^4x \left( {1\over
2}\,\nabla^m\,\hat{F}^B_{ia}\,\nabla_m\,\hat{F}^{aiB} -
2\,\partial^m\,\omega\,\partial_m\,\omega -{\kappa^4\over
4}\,\omega^2\,\xi^2\,A^mA_m \right)~, \label{EH1comp}
\eeqa
where
$$
\xi^2 = \xi^{ik}\xi_{ik}
$$
and
\beq
\nabla^m\,\hat{F}^B_{ai} = \partial^m\,\hat{F}^B_{ai} - {1\over 2}
A^m\,\hat{F}^A_{ai}\,\epsilon^{AB}~. \label{defnabla}
\eeq

Further, the constraint \p{cnstr} allows one to eliminate $\omega(x)$
\beq \label{OMeh}
\omega(x) = {1\over \sqrt{2}\;|\kappa|}\;\sqrt{1 + \frac{\kappa^2}{2}\,(\hat{F}^B\hat{F}^B)} \label{EHom}
\eeq
and from \p{Vexpr} we find $V^{(ik)}_m$
\beq
V^{(ik)}_m = \kappa^2\,\hat{F}^{B(i}_a\partial_m\,F^{Bak)}
= \kappa^2\,{\cal V}_m^{(ik)}~. \label{Veh}
\eeq

Putting all this together, passing to $F^{Bai} =
(\sqrt{2}\,|\kappa|\,\omega)^{-1}\,\hat{F}^{Bai}$ and expressing the auxiliary
gauge field $A_m$ from its algebraic equation of motion
\beq
\delta A_m:
\qquad A_m = 2\, \frac{F^{aiA}\partial_m F^B_{ai}\epsilon^{AB}}{(F^BF^B) -
\kappa^2\,\xi^2}~, \label{Aeh}
\eeq
we obtain the following final form of the physical bosons action
\beqa
S_{QEH} &=& S^{(1)}_{EH} +{\kappa^2\over
2}\;\int d^4x \,{\cal V}^{(ik)}_m(x)\, {\cal V}^m_{(ik)}(x) \nn
&=& {1\over
2}\,\int d^4 x \left\{ \frac{\partial_m\, F^B_{ai}\,\partial^m\,F^{aiB} }{1
- \frac{\kappa^2}{2}\,(F^AF^A)}\, + \kappa^2\;
\frac{(F^B_{ai}F^{iD}_b)\,(\partial_m\,F^{aB}_j\,
\partial^m\,F^{bjD})}{\left[1 - \frac{\kappa^2}{2}\,(F^AF^A)\right]^2}
\right. \nn
&& \left. - \;
\frac{(F^{aiD}\,\partial^m\,F^C_{ai})\,\epsilon^{DC}\,
(F^{bkE}\,\partial_m\,F^H_{bk})\,\epsilon^{EH}}{\left[1 -
\frac{\kappa^2}{2}\,(F^AF^A)\right]\left[(F^BF^B) - \kappa^2\xi^2 \right]} \right\}~.
\label{finEH}
\eeqa
The constraint \p{EHcnstr} rewritten in terms of
$F^{Bai}$ should be also added
\beq
F^{a(iA}\,F^{j)B}_a\,\epsilon^{AB} + \xi^{ij}
= 0\,. \label{EHcnstr1}
\eeq
Note that it is the same as in the HK
case \cite{cf}, \cite{giot}. In what follows, we shall frequently use the
frame in which (cf. \p{frame0})
\beq
\xi^{11} = \xi^{22} = 0~, \quad \xi^{12} = ia~, \quad \xi^2 = 2 a^2~.
\label{frame2}
\eeq

Eqs. \p{finEH} and \p{EHcnstr1} fully specify the QEH target metric. We
shall explicitly present it in the next Section by solving \p{EHcnstr1} in
terms of four independent target space coordinates. Note that \p{EHcnstr1}
itself eliminates 3 out of the original 8 bosonic component. One more degree of
freedom is traded for the residual $U(1)$ gauge symmetry of the action
\p{finEH} which survives in the WZ gauge \p{WZ1}. It is realized by
the transformations
\beqa
&& \delta F^A_{ai} = \varepsilon\,\epsilon^{AB}\,F^B_{ai} +
\varepsilon\,\kappa^2\,\xi_{(i}^{\;\;\;\;k)}\,F^A_{ak}~, \nn
&& \delta A_m = -2\, \partial_m\,\varepsilon~, \quad
\delta V^{(ik)}_m = \kappa^2\,\xi^{(ik)}\,\partial_m\,\varepsilon +
\kappa^2\,\varepsilon \left(\xi^{(il)}V_{l\;m}^{\;k} +
\xi^{(kl)}V_{l\;m}^{\;i} \right)~.
\label{residEH}
\eeqa
It is easy to check that the composite gauge fields \p{Veh}, \p{Aeh}
possess just needed transformation properties (in checking this,
one should take into account the constraint \p{EHcnstr} or \p{EHcnstr1}
and the expression \p{OMeh} for $\omega(x)$). The constraint \p{EHcnstr1}
is invariant in its own right.

Note that the $SU(2)\otimes U(1)$ isometry of QEH metric is
originally realized as the rigid symmetries of the ``master''
gauge-invariant Lagrangian \p{EHmast} corresponding to the following Killing
potentials
\beq \label{EHisom1}
\Lambda^{++}_{ab} = Q^{+A}_a\,Q^{+A}_b~, \qquad \Lambda^{++} =
\gamma\,\xi^{(ik)}q^+_iq^+_k~, \quad \gamma = \bar{\gamma} \neq 0~.
\eeq

It is also worth noting that at the special relation between $\xi^2 = 2a^2$ and
Einstein constant $\kappa$ there emerges an enhancement
of this isometry up to
$SU(2,1)$ (or $SU(3)$, depending on the sign of $\kappa^2 a$).
To see this, it is
convenient to deal with $\hat{Q}^{+aA}$. The supercurrent to which $V^{++}$
couples in \p{EHmast} exhibits an extra rigid symmetry
\beq
\delta
\hat{Q}^{+A}_a = \varepsilon^{Ai}_a\,q^{+k}\,\xi_{ik}~, \quad \delta q^{+i}
= {1\over \kappa^2}\,\varepsilon^{iA}_a\,Q^{+aB}\,\epsilon^{AB}~.
\label{extra}
\eeq
However, the kinetic part of \p{EHmast} is invariant only
providing that the additional ``self-duality'' constraint is imposed
on the transformation parameters:
\beq \epsilon^{AB}\,\varepsilon^{aB}_i =
\kappa^2\,\xi_{ij}\varepsilon^{ajA}~. \label{selfdual}
\eeq
It leaves just $4$ independent parameters in $\varepsilon^{aB}_i$.
For its self-consistency one should require
\beq
{1\over 2}\, \kappa^4\,\xi^2 = \kappa^4\,a^2= 1 \quad \Rightarrow \kappa^2\,a = \pm 1~. \label{enh}
\eeq
It can be shown that under \p{selfdual}, \p{enh} the transformations
\p{extra} close on $SU(2)\times U(1)$ generated by the Killing
potentials \p{EHisom1} and so, together with the latter, they form
the $SU(2,1)$ (or $SU(3)$, depending on the sign of $\kappa^2\,a$)
isometry group. In other words, in this case four physical bosons
parametrize the homogeneous quaternionic space $SU(2,1)/SU(2)\otimes U(1)$
(or $SU(3)/SU(2)\otimes U(1) \sim {\mbb C}P^2$).
This enhancement of isometry is
specific just for the QK case: it does not allow HK limit
in view of the presence of inverse powers of $\kappa^2$ in \p{extra}
\footnote{Actually, a similar enhancement of the isometry group to
$SU(2,1)$ or $SU(3)$ emerges also in the QTN case, once again at
the special relation between $\kappa^2$ and the TN ``mass'' parameter
\cite{{bgio},{iv1}}.}.

Finally, we recall that in the HK case \cite{giot} the HSS action
corresponding to EH metric can be concisely written in terms of one
hypermultiplet superfield which comprises just 4 independent physical bosons
parametrizing the target EH manifold. The same can be done in the QEH case.
The relevant action looks rather complicated due to the presence of
non-trivial dependence on the compensator superfield $q^+_i$. Nevertheless,
such a representation is useful for understanding one important feature of
the resulting QEH metric. Namely, this metric (see next Section) involves as
an independent parameter not $a$ defined in \p{frame2}, but its square
$a^2$.  The substitution $a^2 \rightarrow -a^2$ also yields an admissible
QEH type metric. However, the latter cannot be reproduced using the $SO(2)$
quotient; to recover it within the quotient construction, one should
start with $SO(1,1)$ as the gauge group. We explain this in Appendix B,
specializing for simplicity to the HK case. The same reasoning
equally applies to the QK case modulo $\kappa^2$ corrections.

\setcounter{equation}{0}

\section{Geometrical structure of the QEH metric}
\subsection{Coordinates choice and local structure}
The next step in our analysis will be to solve the constraint \p{EHcnstr1}
and to get from \p{finEH} the metric for some definite choice of coordinates.

To this end we choose for the $\,SU(2)_{\rm susy}\,$ symmetry breaking
triplet the form \p{frame2}
\beq\label{f1}
\xi^{\,11}=\xi^{\,22}=0~,\qq\qq
\xi^{\,12}=ia~.
\eeq
The reality constraint \p{real2} on the triplet
$\,\xi^{\,ij}\,$ implies that the parameter $\,a\,$ has to be real. The
constrained coordinates $\,F^{\,a\, i\, A}\,$ are parametrized in terms
of the Pauli-Gursey spinors, adapted to the expected $\,SU(2)_{\rm PG}\,$
invariance, according to
\beq\label{f2}
F^{\,a\, i=1\, A=1}=\alf(t)\,g^a,\qq F^{\,a\,
i=2\, A=2}=i\be(t)\,g^a~, \qq\qq t=g^a\ol{g}_a\geq 0~,
\eeq
with real functions
$\,\alf(t)\,$ and $\,\be(t)\,.$ The remaining components follow from the
reality conditions \p{real}. As a consequence
\beq\label{f3}
F^{a (i
B}\,F^{b j) A}\eps_{ab}\eps^{BA}+\xi^{(ij)}=0 \qq \Longrightarrow\qq
t\,\alf\be=a/2~.
\eeq
The scalar functions $\,A\,$ and $\,D\,$ defined by
\footnote{We hope that the use of the same characters to denote
the structure functions appearing in the QTN (eqs. \p{QTNmetric}) and QEH
metrics will not lead to any confusion.},
\beq A \equiv 1 -
\frac{\kappa^2}{2}\,(F^BF^B)~, \quad D \equiv (F^BF^B) -\kappa^2\xi^2~,
\label{defAB}
\eeq
become now
$$
A=1-\ka^2\,t(\alf^2+\be^2)~,\qq\qq\qq
D=2\,t(\alf^2+\be^2)-2\ka^2\,a^2~.
$$
To express the metric it is convenient
to use the basis of one-forms such that
\beq\label{f4} \left\{\barr{l} \dst
g^a\,d\ol{g}_a=\frac 12(dt-it\,\si_3)~,\qq\qq g^a\,dg_a=\frac
t2(\si_2+i\si_1)~,\\[4mm] \dst \qq \qq\qq d\si_i=
-\frac 12\eps_{ijk}\,\si_j\wedge\si_k~.\earr\right.
\eeq
For $\ka^2=0\,$, the
identification of the resulting metric with Eguchi-Hanson's \cite{fg}
requires the relation $t(\alf^2+\be^2)=\sqrt{t^2+a^2}\,$, which in turn
implies
\beq\label{f5}
\dst\alf^2=\frac
12\left(\sqrt{1+\frac{a^2}{t^2}}+1\right),\qq\qq \be^2=\frac
12\left(\sqrt{1+\frac{a^2}{t^2}}-1\right).
\eeq
The same choice for
$\ka^2\neq 0$ and the change of variable $\,s=\sqrt{t^2+a^2}\,$ leads to the
following local form of the metric
\beq\label{f6}
\dst g=\frac
1{4C^2}\left[\frac{sB}{s^2-a^2}ds^2+sB\,(\si_1^2+\si_2^2)+
\frac{s^2-a^2}{sB}\,\si_3^2\right],
\eeq
where
\beq\label{f7}
C=1-\ka^2s~,\qq\qq\qq sB= {1\over 2}\,D = s-\ka^2 a^2~.
\eeq
We see that the metric includes the real parameter
$a^2\,$, and not $a$, in accord with the reasoning
adduced at the end of the previous Section and in Appendix B. The
choice of $a^2 < 0$ is equally admissible, it also yields a QK metric.
The same is of course true for $\,\kappa^2\,$ which needs just to be real.

Using this explicit form of the metric we can check its geometrical
structure.
It is convenient to use the vierbein basis for which we have
$$g=\dst\sum_{A=0}^{A=3}e_A^2~.$$
To this end we take
\beq\label{f8}
e_0=\frac 1{2C}\sqrt{\frac{sB}{s^2-a^2}}ds~, \quad
e_3=\frac 1{2C}\sqrt{\frac{s^2-a^2}{sB}}\,\si_3~,\quad
e_1=\frac{\sqrt{sB}}{2C}\,\si_1~,\quad
e_2=\frac{\sqrt{sB}}{2C}\,\si_2~.\eeq
The spin connection $\,\om_{AB}\,$ and the curvature $\,R_{AB}\,$ are
defined \cite{egh} by
$$de_A+\om_{AB}\wedge e_B=0~,\qq\qq R_{AB}=d\om_{AB}+\om_{AC}\wedge\om_{CB}~,
\qq A,B,C=0,1,2,3~.$$
The self-dual components are
$$R^{\pm}_i=R_{0i}\pm\frac 12\,\eps_{ijk}\,R_{jk}~,\qq\qq
\la_i^{\pm}=e_0\wedge e_i-\frac 12\,\eps_{ijk}\, e_j\wedge e_k~,
\qq i,j,k=1,2,3,$$
and similarly for the Weyl tensor.
One then defines the matrices ${\cal A},{\cal B}$ and ${\cal C}$ by
\beq\label{f9}
\left(\barr{cc} R^+ \\[3mm] R^-\earr\right)=
\left(\barr{cc} {\cal A} & {\cal B}^t\\[3mm] {\cal B} & {\cal C} \earr\right)
\left(\barr{cc} \la^+ \\[3mm] \la^-\earr\right),\eeq
from which we deduce the scalar curvature and Weyl tensor by
$$\dst \frac R4=\Lambda={\rm tr}\,{\cal A}=\,{\rm tr}\,{\cal C}~,\qq\qq
W^+={\cal A}-\frac{{\rm tr}\,{\cal A}}{3}\,{\mbb I}~,
\qq\qq W^-={\cal C}-\frac{{\rm tr}\,{\cal C}}{3}\,{\mbb I}~.$$
Easy computations give
\beq\label{f10}
\dst {\cal B}=0~, \qq \Lambda=-12\ka^2~,\qq W^+=0~,\qq W^-=
-4a^2\frac{C^2}{(sB)^3}\ {\rm diag}\,(1,1,-2)~.
\eeq
The vanishing of $\,{\cal B}\,$ implies \cite{egh} that the metric (\ref{f7})
is Einstein, and its Weyl tensor is anti-self-dual, as expected.

\subsection{The hyperbolic monopole structure}
Obviously, when the Einstein constant $\kappa^2\to 0\,$, the metric (\ref{f6})
reduces to Eguchi-Hanson's \cite{eh} which is a particular case of the
multicentre
metrics \cite{egh}. They can be written
\beq\label{f11}
\frac 1V(d\tau+A)^2+V\,h~,\eeq
where  $A$ is some connection 1-form, $\,h\,$ the flat 3-space metric while
$V$ and $A$ are related by the monopole equation
\newcommand{\dec}{\mathop{\smash{\star}}}
$$\dec_{h}\,dV=dA~.$$
As shown by Pedersen in \cite{pd}, as far as $\,a^2>0\,,$ the monopole
structure, properly generalized,
survives at the level of the quaternionic extension (\ref{f6}). Indeed it can
be written, using angular coordinates, as follows
$$\frac 1{4C^2}\left\{\frac{s^2-a^2}{sB}(d\psi+\sin\tht d\phi)^2+
sB\left[\frac{ds^2}{s^2-a^2}+d\tht^2+\sin\tht^2 d\phi^2\right]\right\}.$$
The substitution $\,s/a=\coth\rho\,$ gives
\beq\label{f12}
\frac a{4(\sinh\rho-\kappa^2 a\cosh\rho)^2}\left\{\frac 1V (d\psi+A)^2+
V\gamma({\mathbb H}_3)\right\},\eeq
with
\beq\label{monop1}
V=\coth\rho-\ka^2 a~,\qq A=\cos\tht\, d\phi~,\qq \gamma({\mathbb H}_3)=
d\rho^2+\sinh^2\rho\left(d\tht^2+\sin\tht^2 d\phi^2\right)~,\eeq
where $\gamma({\mathbb H}_3)$ is the metric of hyperbolic 3-space. The
monopole equation reads now
\beq\label{monop2}
\dec_{\ga}\,dV=dA~.\eeq
The quaternionic extension of Eguchi-Hanson retains some structure from its
hyperk\" ahler origin through the monopole equation, the main difference
being that flat 3-space is replaced by hyperbolic 3-space.

The situation is somewhat different for $\,a^2<0\,$ since then we have to
use the change of variables $\,s/a=1/\sinh\rho\,$ to recover the hyperbolic
3-space. The metric can be written as
\beq\label{ff12}
\frac{a\cosh^2\rho}{4(\sinh\rho-\kappa^2a)^2}\left\{
\frac 1V\, (d\psi+A)^2+
V\,\tilde{\gamma}({\mathbb H}_3)\right\},\eeq
with
\beq\label{monop3}
V=\frac 1{\sinh\rho}+\kappa^2a~,\qq\qq A=\cos\tht\, d\phi~, \qq\qq
\tilde{\gamma}=\frac{\gamma({\mathbb H}_3)}{\cosh^2\rho}~.\eeq
The monopole equation is now
\beq\label{monop4}
\dec_{\tilde{\ga}}\,dV=dA~.\eeq

\subsection{Global structure : the complete metrics}
Before considering generic values of the real parameters $\,\ka^2\,$
and $\,a^2\,$, let us examine separately some special cases.

First, for $\,a^2\to 0\,$ the metric (\ref{f6}) simplifies to
$$g=\frac 1{(1-\kappa^2 s)^2}\left[\frac{ds^2}{4s}+
\frac s4(\si_1^2+\si_2^2+\si_3^2)\right],$$
in which we recognize the conformally flat structure of the spere $\,S^4\,$
for $\,\kappa^2>0\,$, of the flat space for $\,\kappa^2=0\,$, and of the hyperbolic
space $\,H^4\,$ for $\,\kappa^2<0\,.$

Second, when $\,sB\,$ and $\,A\,$ are proportional to each other, the Riemann tensor in the
vierbein basis has constant components, showing that we deal with a symmetric
space. This can happen only for $\,a^2>0\,.$  Defining $\,\ka^2a=\eps=\pm 1\,$
gives $\,sB=-\eps a C\,.$ In this cases the 2-form
\beq\label{g1}
\Omega^{(+)}=e_0\wedge e_3+e_1\wedge e_2\equiv \frac 12(I^+)_{\mu\nu}\,
dx^{\mu}\wedge dx^{\nu}\eeq
is closed. One can check that $\,(I^+)_{\mu}^{~\nu}\,$ defines
a true complex
structure and therefore the metric (\ref{f6}) is also k\" ahlerian.
It must be
either $\,\dst{\mbb C}P^2\sim SU(3)/U(2)\,$ or its complete (albeit
non-compact) dual $\,\widetilde{{\mbb C}P}^2\sim SU(2,1)/U(2)\,.$ To check
this at the level of the metric we write it in the form
\beq\label{g2}
\frac{a^2}{4(s-\eps a)^2}\left\{\frac{ds^2}{s+\eps a}+
(s-\eps a)(\si_1^2+\si_2^2)+(s+\eps a)\si_3^2\right\}.\eeq
Simplifying the $\,a^2\,$ factor and the change of variable
$$\dst \frac 1{s-\eps a}=\frac{t^2}{1+\la t^2/6}~,\qq\qq  \la/6=-2\eps a~,$$
bring the metric (\ref{f10}) to the form
\beq\label{g3}
\dst \frac{dt^2}{\dst (1+\la t^2/6)^2}+
\frac{t^2}{\dst 1+\la t^2/6}(\si_1^2+\si_2^2)+
\frac{t^2}{\dst (1+\la t^2/6)^2}\si_3^2~,
\eeq
which is the standard form \cite[p. 384]{egh} of $\,{\mbb C}P^2\,$
for $\,\eps=-1\,$ and its non-compact dual for $\,\eps=+1\,.$ These choices
just amount to the isometry enhancement relation  \p{enh}.

 From now on we will exclude the previous special cases from the analysis
and examine the global properties of the QEH metric
\beq\label{g4}
\frac 1{4(1-\ka^2s)^2}\left\{\frac{s-\ka^2a^2}{s^2-a^2}ds^2+
(s-\ka^2a^2)(\si_1^2+\si_2^2)+\frac{s^2-a^2}{s-\ka^2a^2}\si_3^2\right\}.\eeq
As we will see there are several complete (but non-compact) metrics which
we now enumerate :
\brm
\item[$\bullet$] First case : $\quad \ka^2<s<\nf\,.$
Let us first observe that the metric positivity requires, for
$\,a^2>0\,,$ the additional constraint $\,\kappa^2a<1\,.$
For large $\,s\,$ the metric is proportional to
$$ g\approx d\tau^2+\frac{\tau^2}{4}(\si_1^2+\si_2^2+\si_3^2)~,
\qq\qq \tau=1/\sqrt{s}~,$$
a typical nut behavior in the terminology of \cite{gh}. This means that
$\,\tau=0\,$ is an apparent singularity which can be removed by going back to
cartesian coordinates and the metric is smooth there.

The situation is fairly different at $\,s=1/\ka^2\,$ where the metric behaves like
\beq\label{g5}
\frac{(1-\ka^4a^2)}{4\ka^2(1-\ka^2s)^2}\left\{\frac{\ka^4ds^2}{1-\ka^4a^2}+
\si_1^2+\si_2^2+\frac{\si_3^2}{1-\ka^4a^2}\right\}.
\eeq
However this singularity is ``infinitely far" since we have
$$\dst \int^{1/\ka^2}\ \frac{ds}{1-\ka^2s}=\nf~,$$
so it does not jeopardize completeness. We recognize as
conformal structure,
for fixed $\,s\,$, the Berger (or squashed) sphere $\,S^3\,$ with the metric
\beq\label{berger}
\si_1^2+\si_2^2+I\si_3^2~,\qq\qq I=\frac 1{1-\ka^4a^2}~,\eeq
where the constant $\,I\,$ is bigger or
smaller than 1 according to the sign of $\,a^2\,.$
If we set
$$\rho^2=\frac 1{\ka^2 s}\ \in[0,1)~,\qq\qq m^2=-\ka^4a^2~,$$
the metric (\ref{g4}) becomes
\beq\label{g6}
\frac 1{\kappa^2}
\frac 1{(1-\rho^2)^2}\left\{\frac{1+m^2\rho^2}{1+m^2\rho^4}\,d\rho^2+
\rho^2(1+m^2\rho^2)\frac{(\si_1^2+\si_2^2)}{4}+
\rho^2\,\frac{1+m^2\rho^4}{1+m^2\rho^2}\,\frac{\si_3^2}{4}\right\},\eeq
which is, for $\,\kappa^2>0\,,$ and for any real $\,a^2\,,$ the complete metric
given by Pedersen in \cite{pd}.

\item[$\bullet$] Second case : $a\leq s<1/\ka^2~.$\erm

\noindent This complete metric does exist only for $\,a^2>0\,.$ Positivity requires
the additional constraint $\,\kappa^2a<1\,.$

Near to the singularity $\,s=a\,$ the metric behaves as
\beq\label{g7}
g\approx \frac 1{2(1-\ka^2a)}\left\{d\tau^2+\frac{\tau^2}{(1-\ka^2 a)^2}\,
\si_3^2
+\frac a2\,\left(\si_1^2+\si_2^2\right)\right\},\qq \tau=\sqrt{s-a}~.\eeq
Using angular coordinates, one has
$$\si_1^2+\si_2^2=d\tht^2+\sin^2\tht d\phi^2~,
\quad \si_3=d\psi+\cos\tht d\phi~,
\quad\tht\in[0,\pi],\quad \phi\in[0,2\pi]~,\quad\psi\in[0,4\pi]~.$$
So we can impose the constraint
\beq\label{g8}
\frac 1{1-\ka^2 a}=k~,\qq k=2,3,\ldots\qq\Longrightarrow\qq
\ka^2 a=\frac{k-1}{k}<1~,\eeq
and take for $\,\psi\,$ the interval of variation $[0,4\pi/k]\,.$
Then the metric can be smoothly continued through $\,s=a\,$ which is a bolt
\cite{gh} of twist $\,k\,.$

Locally, the boundary at $\,s=1/\ka^2\,$ has the same  structure (Berger
metric) as in the first case considered previously. However, globally, the
identification of $\,\psi\,$ under the action of $\,{\mbb Z}_k\,$ gives for
the conformal structure the Lens space $\,S^3/{\mbb Z}_k\,.$

Let us point out that our conclusions are in complete agreement with
the full classification of self-dual Einstein metrics given by Hitchin
in \cite{hi}. One will find in \cite{ma} an interesting discussion of
various Einstein extensions of Eguchi-Hanson's metrics in different
coordinates. Beyond the quaternionic and the non-quaternionic ones
there is a K\" ahler-Einstein extension, first given in \cite{gp}, whose
completeness was discussed in \cite{pn}.

\subsection{K\" ahler metrics in the conformal class}
It was first proved in \cite{pp} that the conformal class of Pedersen
metric contains Le Brun metric, which can be written
\beq\label{lb1}
\frac{dr^2}{W}+\frac{r^2}{4}(\si_1^2+\si_2^2)+\frac{r^2}{4}\,W\si_3^2~,\qq
W=1+\frac{A}{r^2}+\frac{B}{r^4}~.\eeq
It is K\" ahler and scalar-flat, and for $\,A=0\,$ it reduces to
Eguchi-Hanson. Le Brun considered the special choice of parameters
\beq\label{lb2}
A=(n-1)~,\qq\qq B=-n~,\qq n=0,1,2,\ldots\qq\quad r\geq 1~,\eeq
for which he proved completeness.

It follows that the QEH metric considered here should be (locally)
conformal to (\ref{lb1}). Imposing the K\" ahler constraint gives for
the conformal factor $\,\rho=C^2,$ and for the resulting metric
\beq\label{conforme}
\hat{g}=\frac{(s-\kappa^2a^2)}{4(s^2-a^2)}ds^2
+\frac{(s-\kappa^2a^2)}{4}(\si_1^2+\si_2^2)+
\frac{(s^2-a^2)}{4(s-\kappa^2a^2)}\si_3^2~.\eeq
Its identification with the Le Brun metric is performed through
the relations
\beq\label{lb3}
r^2=s-\kappa^2a^2~,\quad W=\frac{s^2-a^2}{(s-\kappa^2a^2)^2}~,\quad
A=2\kappa^2a^2~,\quad B=a^2(\kappa^4a^2-1)~.\eeq

An important observation is that, in our notations, its Weyl tensor which is
self-dual,  $\,W^{-}=0\,,$ and its complex structure given by
$$\Omega=ds\wedge\si_3-sB\,\si_1\wedge\si_2~,$$ have the opposite
orientation. However, in the same conformal class, we can find another K\"
ahler metric, whose complex structure and Weyl tensor have the {\em same}
orientation. A local computation gives for it \beq\label{new}
\frac{dr^2}{V}+\frac{r^2}{4}(\si_1^2+\si_2^2)+
\frac{r^2}{4}\,{V}\,\si_3^2~,\qq {V}=\frac 1{r^4}+\frac{A}{r^2}+B~,\eeq that
is indeed a close cousin of (\ref{lb1}) ! This shows that the Le Brun metric
with $\,B=1\,$ has two commuting complex structures of opposite self-duality
and is scalar-flat. Its new partner is still  K\" ahler but it has a non-constant
scalar curvature  $\,R=-8(B-1)/r^2$. Complete metrics are given, for
instance, by \beq\label{newcomplete} A=-n-2~,\qq B=n+1~,\qq r\geq 1~,\qq
n=1,2,\ldots\eeq where $\,r=1\,$ is a bolt of twist $\,n\,$ and the metric
is locally asymptotically flat at infinity. The scalar curvature is
negative.

\section{Conclusions}
In this paper, starting from the most general HSS action of the coupled
$N=2$ SG-hypermultiplets system, we derived the most general bosonic QK
sigma model action which is a useful tool for the explicit
local construction of the metric on an arbitrary $4n$ dimensional
QK manifold. We found that this action admits two flat limits. One of them
yields a general HK sigma model action with $4n$ dimensional target,
while the second gives rise to the HK sigma model with $4(n+1)$
dimensional target
space corresponding to the general superconformally-invariant
self-coupling of $(n+1)$
rigid hypermultiplets. We worked out the HSS version of the QK quotient
approach and applied it to give a new derivation of the QK extensions of the
four-dimensional
Taub-NUT and Eguchi-Hanson metrics. We studied
in detail local and global properties of the
QEH metric and compared it with various examples
of self-dual Einstein metrics known in
literature. We showed that the HSS formulation allows one
to readily reveal the enhancement of the $SU(2)\otimes U(1)$
isometry of the QEH metric to $SU(3)$
or $SU(1,2)$ at the special ratio of its ``mass''
parameter and the Einstein constant $\kappa^2$ (or $Sp(1)$ curvature).

As possible directions of further study let us mention the explicit
construction of metrics on QK analogs of the higher-dimensional toric
HK manifolds, as well as four-dimensional QK metrics properly generalizing
the multicentre ansatz for HK instantons \cite{gh1}. The HK metrics to be
generalized have a nice description in the HSS approach, both within the
purely geometric setting of ref. \cite{gios2} and
in the lagrangian framework of
refs. \cite{gios1,val1,gov}. It is interesting to see how the relevant HSS
actions are extended to the QK case, and how unique such extensions are.
Being armed with the general HSS action for QK sigma models and
the appropriate convenient quotient construction, we expect
these problems to be amenable for solving. An interesting separate
problem is to find out possible relationships of the QK
sigma models in the HSS formulation with the brane-like solutions of
higher-dimensional supergravities \cite{st}, as well as with the theory
of intersecting branes (e.g., along the lines of ref. \cite{ggpt}).

\section*{Acknowledgments}
E.I. thanks E. Sokatchev and P. Townsend for interest in the work and discussions,
and Directorate of Laboratoire de Physique Th\'eorique
et des Hautes Energies, Universit\'e Paris VII, for the hospitality
extended to him during the course of this work. His work was
accomplished under the Project PAST-RI 99/01 and supported in part
by the RFBR-CNRS Grant No. 98-02-22034, RFBR Grant No. 99-02-18417,
Nato Grant No. PST.CLG 974874 and INTAS Grants INTAS-96-0538,
INTAS-96-0308.

\setcounter{equation}0
\def\theequation{A.\arabic{equation}}
\section*{Appendix A}
Here we show that the fields $A_{\mu\dot\mu}$ and $t^{(\mu\nu)}$
defined in \p{Hmu} do not contribute to the structure of the QK sigma model
action of the fields $F^{ri}(x)$.

It is clear from the $\theta$ expansion \p{thetaexp1} that $t^{(\mu\nu)}$
does not appear in the bosonic part of the action \p{Sq}.
The same is true for the real part of $A^{\dot\mu\mu}$: it is the gauge
field for the local $\gamma_5$ transformations and so cannot couple
to $f^{ai}(x), F^{ri}(x)$ which are $\gamma_5$-neutral.
As for the imaginary part of $A^{\dot\mu\mu}$,
the gauge field for the local scale transformations,
its decoupling is not immediately obvious from \p{Hmu}, \p{Sq}. The
direct computation yields the following general expression for
the relevant piece in the bosonic lagrangian (the precise normalization of
$S^m \sim \mbox{Im} A^m $ is not crucial for our purposes)
\beqa
&& {\cal L}_{S} = S^m(x) J_m(x)~, \nn
&&J_m(x) =
\int du \left(-f^+_a(x,u)\,A^{-a}_m(x,u) +\kappa^2\,(u^-f^{+})^2
\,F^+_r(x,u)\,B^{-r}_m(x,u) \right)~. \label{Bcoupl}
\eeqa
However, ${\cal L}_{S} \sim S_m S^m$ on the shell of eqs. \p{feq}, \p{Feq},
\p{cnstr}, so $S^m$ decouples as well, in accord with the fact that the
fields $F^{ri}(x)$ have zero conformal weight. This can be shown in the case of
arbitrary $L^{+4}(Q, v^+, u^-)$. For simplicity, just to give the idea of
the proof, we specialize to the ``maximally flat''
$Sp(1,n)/Sp(1)\otimes Sp(n)$ case with $L^{+4} = 0$.

It is convenient to deal with the superfields $\hat{Q}^{+}_r =
|\kappa|\,(u^-q^+)\,Q^+_r$. The current $J_m(x)$ equals to
\beq
J_m(x) = \int du \left(-f^+_a\,A^{-a}_m + \hat{F}^+_r\,\hat{B}^{-r}_m
\right)~. \label{curr1}
\eeq
Using identities
$$
f^{+a} = \partial^0\, f^{+a} = [\,\partial^{++}, \,\partial^{--}\,]\, f^{+a}~, \quad
\hat{F}^{+r} =  \partial^0\, \hat{F}^{+r} = [\,\partial^{++}, \,\partial^{--}\,]\,
\hat{F}^{+r}~, \quad
$$
and integrating by parts, one can rewrite \p{curr1} as
\beqa
J_m(x) &=& \int du \left(\partial^{--}f^+_a\,\partial^{++}A^{-a}_m -
\partial^{++}f^+_a\,\partial^{--}A^{-a}_m  \right. \nn
&& \left. - \;
\partial^{--}\hat{F}^+_r\,\partial^{++}\hat{B}^{-r}_m
+ \partial^{++}\hat{F}^+_r\,\partial^{--}\hat{B}^{-r}_m\right)~. \label{curr2}
\eeqa
Further, eqs. \p{feq}, \p{Feq} for $f^{+a}, \hat{F}^{+r}$ and the non-dynamical
harmonic equations of motion for $A^{-a}_m, \;\hat{B}^{-r}_m$ read in this case
\beqa
&& \partial^{++}f^{+a} = \partial^{++}\hat{F}^{+r} = 0~, \nn
&& \partial^{++}A^{-a}_m = 2\, (\partial_m - S_m)f^{+a}~, \quad
\partial^{++}\hat{B}^{-r}_m = 2\, (\partial_m - S_m)\hat{F}^{+r}~.
\eeqa
Using them, one reduces \p{curr2} to
\beq
J_m(x) = (\partial_m - S_m) \int du \left(f^{+a}\partial^{--}f^{+}_a -
\hat{F}^{+r}\,\partial^{--}\hat{F}^+_r \right)~. \label{curr3}
\eeq
The expression on which $(\partial_m - S_m)$ acts is just the l.h.s. of the
constraint \p{cnstr}, so
\beq
J_m(x) = -{1\over \kappa^2}\, S_m(x) \quad \Rightarrow \quad {\cal L}_{S} =
-{1\over \kappa^2}\, S_m(x)S^m(x)~. \label{curr4}
\eeq

It is straightforward, though somewhat tiresome, to show that in the general case
$J_m(x)$ and ${\cal L}_{S}$ are given by the same expressions \p{curr3},
\p{curr4}. One should use the general $f^+, F^+$ equations \p{feq}, \p{Feq} and the
corresponding harmonic equations for $A^{-a}_m, \;B^{-r}_m$.

\setcounter{equation}0
\def\theequation{B.\arabic{equation}}
\section*{Appendix B}
Here, on the simple example of the standard hyper-K\"ahler EH metric in the
HSS approach \cite{giot}, we explain the origin of the freedom in  changing
the sign of the EH ``mass'' parameter $a^2$. The same reasoning applies to the
QEH case.

In the HK limit the QEH Lagrangian \p{EHmast} becomes \cite{giot}
\beq
{\cal L}^{+4}_{EH} =
Q^{+B}_a D^{++} Q^{+aB} +
V^{++}\left(\epsilon^{AB}Q^{+aA}Q^{+B}_a
+ \xi^{(ik)}u^+_iu^+_k \right)~. \label{EHhyperk}
\eeq
Passing to the complex combinations
$$
Q^+_a = Q^{+1}_a + iQ^{+2}_a~, \quad \bar Q^+_a = Q^{+1}_a -i Q^{+2}~,
$$
making the field redefinition
$$
Q^+_a = u^+_a\, {1\over \sqrt{2}}\,\omega + u^-_a \,L^{++}~, \quad
\bar Q^+_a = u^+_a\,{1\over \sqrt{2}}\, \bar\omega + u^-_a \, \bar
L^{++}~,
$$
and, finally, eliminating the non-propagating superfields $L^{++}$, $\bar L^{++}$
from \p{EHhyperk} by their algebraic equations of motion, we equivalently
rewrite \p{EHhyperk} as
\beq
{\cal L}^{+4}_{EH} = -{1\over 2}\; (D^{++} - iV^{++})\,\omega\,(D^{++} +
iV^{++})\,\bar\omega + V^{++}\,\xi^{++}~. \label{omega}
\eeq
The initial gauge group now acts as $U(1)$ phase transformations of $\omega$,
$\bar{\omega}$. Choosing the gauge
$$
\omega = \bar\omega~,
$$
and eliminating $V^{++}$ by its equation of motion
$$
V^{++} = \frac{\xi^{++}}{\omega^2}~,
$$
we obtain, for the HSS lagrangian corresponding to EH
metric, the representation in terms of one hypermultiplet $\omega$ \cite{giot}
\beq
{\cal L}^{+4}_{EH} = -{1\over 2}\left[ (D^{++}\omega)^2 -
\frac{(\xi^{++})^2}{\omega^2} \right]~. \label{omegaEH}
\eeq

 From this form of the action it immediately follows that the relevant HK
metric in the fixed frame \p{frame2} can depend only on $a^2$.
Further, the action obtained from \p{omegaEH} by changing the sign of
the ``potential'' term, or, equivalently, by the substitution $a^2 \rightarrow -a^2$,
also yields some HK metric in the bosonic sector. These two metrics are related via
the replacement $a^2 \rightarrow - a^2$. However, the second metric {\it cannot}
be recovered from the above
$SO(2)$ quotient construction. Indeed, this could be achieved only at cost of the change
$a \rightarrow ia$, which would ruin the reality of the actions \p{EHhyperk}, \p{omega}.

Nevertheless, \p{omegaEH} with the ``wrong'' sign of the second term can
be inferred from a proper quotient construction.
One should start from the gauge-invariant action of
two {\it real} $\omega$ hypermultiplets with $SO(1,1)$ as the gauge group:
\beq
{\cal L}^{+4}_{EH}{}' = -{1\over 2}\; (D^{++} - V^{++})\,\omega_1\,(D^{++} +
V^{++})\,\omega_2 + V^{++}\,\xi^{++}~. \label{omega0}
\eeq
Here $\xi^{++}$ has the same reality properties as in the previous case, $\omega_{1,2}$
undergo scale transformations with the real
analytic parameter $\varepsilon\,(\zeta)$:
\beq
\delta \omega_1 = \varepsilon\,\omega_1~, \quad  \delta \omega_2 =
-\varepsilon\, \omega_2~.
\eeq
The gauge superfield $V^{++}$ transforms in the same way as in the $SO(2)$ case, $\delta
V^{++} = D^{++}\varepsilon $. Choosing the gauge
$$
\omega_1 = \omega_2 \equiv \omega
$$
and eliminating $V^{++}$,
$$
V^{++} = -\frac{\xi^{++}}{\omega^2}~,
$$
one obtains the desirable lagrangian
\beq
{\cal L}^{+4}_{EH}{}' = -{1\over 2}\left[ (D^{++}\omega)^2 +
\frac{(\xi^{++})^2}{\omega^2} \right]~. \label{1omegaEH}
\eeq
related to \p{omegaEH} just through the change $a^2 \rightarrow -a^2$.

A similar mechanism works in the QK case, demonstrating the
existence of the two types of QEH metrics related by the same change
of the parameter $a^2$. The QK analogs of the actions \p{omegaEH},
\p{1omegaEH} look rather complicated as they involve extra terms
containing the compensator superfield $q^+$. However, the $SU(2)$ breaking
parameters still appear as $(\xi^{++})^2 \sim a^2$.

\end{document}